\documentclass[]{IEEEtran}
\IEEEoverridecommandlockouts
\usepackage{cite}
\usepackage{amssymb,amsfonts}
\usepackage{algorithmic}
\usepackage{graphicx}
\usepackage{textcomp}
\usepackage{xcolor}
\usepackage{booktabs}
\usepackage{pgfplots}
\usepackage{float}
\usepackage{stfloats}
\usepackage{tikz}
\usepackage{makecell}
\usepackage[linesnumbered,ruled]{algorithm2e}
\usepackage{CJKutf8}
\usepackage{amssymb}
\usepackage[tbtags]{amsmath}
\usepackage[thmmarks,amsmath]{ntheorem} 
\usepackage{svg}
\usepackage{makecell}

\newtheorem{theorem}{Theorem}
\newtheorem{lemma}{Lemma}
\newtheorem{corollary}{Corollary}
\newtheorem{remark}{Remark}

\allowdisplaybreaks

\newcommand*{\circled}[1]{\lower.7ex\hbox{\tikz\draw (0pt, 0pt)%
    circle (.5em) node {\makebox[1em][c]{\small #1}};}}
    
 \qedsymbol{\ensuremath{\square}} 
{   
   \theoremstyle{nonumberplain}
   \theoremheaderfont{\bfseries}
   \theorembodyfont{\normalfont}
   \theoremsymbol{\ensuremath{\square}} 
   
}

\def\BibTeX{{\rm B\kern-.05em{\sc i\kern-.025em b}\kern-.08em
    T\kern-.1667em\lower.7ex\hbox{E}\kern-.125emX}}
\begin{document}

\title{\LARGE A Framework of Arithmetic-Level Variable Precision Computing for In-Memory Architecture: Case Study in MIMO Signal Processing}

\author{
Kaixuan Bao, \textit{Graduate Student Member, IEEE}, Wei Xu, \textit{Fellow, IEEE}, 

Xiaohu You, \textit{Fellow, IEEE}, and Derrick Wing Kwan Ng, \textit{Fellow, IEEE}
\thanks{Part of this paper was presented at the IEEE International Conference on Communication Technology, Nanjing, China, April 2022 \cite{10073086}.

Kaixuan Bao, Wei Xu, and Xiaohu You are with the National Mobile Communications Research Laboratory (NCRL), Southeast University, Nanjing 210096, China (e-mail: kxbao@seu.edu.cn; wxu@seu.edu.cn; xhyu@seu.edu.cn). 

Derrick Wing Kwan Ng is with the School of Electrical Engineering and Telecommunications, The University of New South Wales, Sydney, NSW 2052, Australia (e-mail: w.k.ng@unsw.edu.au). 
}
}

\maketitle

\begin{abstract}
Computational complexity poses a significant challenge in wireless communication. Most existing attempts aim to reduce it through algorithm-specific approaches. However, the precision of computing, which directly relates to both computing performance and computational complexity, is a dimension that is fundamental but rarely explored in the literature. With the emerging architecture of in-memory computing, variable precision computing (VPC) is enabled, allowing each arithmetic operation to be processed with a distinct and specifically optimized computing precision. In this paper, we establish a unified framework of arithmetic-level variable precision computing (AL-VPC), which aims to determine the optimized computing precision for each arithmetic operation. We first develop an arithmetic propagation error model exploiting stochastic analysis, and then formulate a mathematical optimization problem to strike balance between computing performance and computational complexity. Two algorithms, namely, offline VPC and online VPC, are proposed to solve the problem considering various practical concerns. Particularly, in a case study on zero-forcing (ZF) precoding, we reveal the Pareto boundary between computing performance and complexity, which exhibits up to a 60\% sum-rate enhancement or equivalently up to a 30\% complexity reduction compared to the traditional fixed-length methods. 
\end{abstract}


\setlength{\abovedisplayskip}{6pt} 
\setlength{\belowdisplayskip}{6pt}

\section{Introduction}
Signal processing for massive multiple-input multiple-output (MIMO) in future wireless communication networks is confronted with new challenges related to its required computational complexity\cite{10024766,10623856}, stemming from its extensive matrix operations and the demand for high data rates \cite{8907829}. Previous research endeavors in MIMO signal processing mainly focused on designing particular algorithms to mitigate computational complexity, e.g., by exercising meticulous control over the steps of iterations\cite{10400549,9112345}. Since these approaches are algorithm-specific, they lack generality.

In recent years, a novel hardware architecture, known as in-memory computing, has emerged as a promising solution aimed at substantially reducing computational latency from a hardware perspective\cite{MUTLU201928}. This architecture exhibits a high level of generality that it could benefit all signal processing algorithms\cite{9626555}. Unlike the traditional Von Neumann architecture, which stores data in memory and processes data in a central processing unit (CPU), the in-memory computing architecture performs both data storage and processing within the memory. This innovation effectively mitigates the predominant latency stemming from data transfer between the CPU and memory. In general, in-memory computing can be classified into two categories: process in memory (PIM)\cite{8686556} and process near memory (PNM)\cite{2019ComputeDRAM}. However,  the intrinsic constraints of the in-memory architecture limits its operation to low bit-width\cite{MUTLU201928}, making it only suitable for low-precision calculations. This constraint often results in computational errors that significantly undermine the computing performance of signal processing\cite{10226413,9626555}. Indeed, this inherent limitation of in-memory computing naturally calls for investigation into the relationship between computing precision, latency, and the performance.

Signal processing algorithms for MIMO communication, e.g., multiuser precoding calculations, inherently demand a high level of computing precision\cite{4610935}. Nevertheless, this pursuit of high precision introduces a conundrum in the form of elevated computational complexity, leading to prolonged computing latency. Since latency is a pivotal performance indicator in future wireless designs, especially for latency-sensitive applications, it becomes crucial to manage the complexity of precoding calculations while satisfying latency requirements\cite{10599869,Xu2023,9007754}. Typical approaches for controlling computational complexity encompass meticulous control over algorithm iterations\cite{7264975}, as well as formulating optimization problems that strike a balance between performance and complexity\cite{9559759,10488685}. There are also some results which implement low bit width computing in a certain part of MIMO signal processing. Typically, in quantized precoding, finite-alphabet equalization has been proposed to reduce the computational complexity of the MMSE decoding process \cite{9110827,7967843,7458830}, achieving near-optimal performance with significantly reduced complexity. 
    
In practical scenarios, computing precision is directly associated with complexity. Essentially, for a given algorithm, a constraint on complexity translates into a constraint on the average computing precision. Under such constraints, one potential approach to enhance the computing performance involves allocating higher computing precision to more crucial steps\cite{678521,10066526}. This approach is commonly known as variable precision computing (VPC)\cite{10.1145/1236463.1236468}. A typical VPC method divides an algorithm into several partitions and subjectively assigns varying computing precision to different partitions\cite{BABOULIN20092526,9739768,8660569,8877436}. However, signal processing algorithms for MIMO communication often pose challenges in terms of separability, making it difficult to directly apply VPC to wireless communication-related algorithms. In deed, even if an algorithm can be logically partitioned, determining which section deserves higher computing precision remains an open problem.

To fully explore the potential of VPC, it is imperative to scale down each partition of the algorithm to its smallest possible unit, with each partition comprising at least a single basic arithmetic operation. This concept is termed arithmetic-level VPC (AL-VPC). However, it is challenging to achieve AL-VPC in practice as it is nearly impossible to subjectively determine the relative importance of each arithmetic operation in an algorithm, particularly within complex MIMO signal processing algorithms such as precoding calculations. Since the importance of each arithmetic operation correlates directly to its error propagation behaviour\cite{probabilistic}, a thorough investigation into this matter becomes imperative. Though this topic has been studied over decades, e.g., \cite{9048893,10.2307/2949434,Tienari1970ASM}, they either base on Gaussian hypothesis or only consider a single arithmetic operation. To the best of our knowledge, there are no comprehensive results that fully unveil the statistical characteristic of error propagation of all basic arithmetics that motivates the study of this work.

In this paper, we establish the first framework of AL-VPC, achieved through the derivation of stochastic arithmetic propagation errors. This approach offers an objective means to assess the significance of each arithmetic operation within an algorithm, consequently enabling the proposal of tailored computing precision assignments for each arithmetic operation. The contributions of this paper are summarized as follows.

\begin{itemize}
\item A general framework of AL-VPC with mathematical modeling is introduced. To model the procedure of a generic algorithm, the concept of expression tree is proposed. Utilizing the expression tree, an optimization problem is formulated to effectively optimize the assignments of computing precision at the arithmetic level.
\item  A stochastic model of arithmetic propagation error is established. The expectation and variance of the error of basic arithmetic operations are derived.
\item Two algorithms, named online VPC and offline VPC, are proposed to solve the optimization problem. Specifically, the online VPC aims to achieve an optimal balance between computing performance and computational complexity, while the offline VPC focuses on facilitating simpler implementation.
\item A case study of multiuser MIMO precoding is examined. The proposed AL-VPC achieves up to a 60\% sum-rate enhancement or equivalently a 30\% complexity reduction, compared to conventional fixed-length computing.
\end{itemize}

The rest of this paper is organized as follows. Section~\uppercase\expandafter{\romannumeral2} introduces the framework of AL-VPC and the related optimization problem. A stochastic model of error propagation is established in Section~\uppercase\expandafter{\romannumeral3}. The solution to the formulated problem is solved in Section~\uppercase\expandafter{\romannumeral4}. Numerical simulations, along with the case study, are presented in Section~\uppercase\expandafter{\romannumeral5}. Section~\uppercase\expandafter{\romannumeral6} concludes this paper.

\section{Arithmetic-Level Variable Precision Computing for Signal Processing}
In this section, we present an introduction to VPC and propose a general framework of AL-VPC. The conventional approach of VPC involves the partitioning of an algorithm into several larger partitions, each comprising hundreds or potentially millions of arithmetic operations \cite{BABOULIN20092526}. These partitions are then assigned with varying levels of computing precision based on subjective assessments of their functionalities. This is suitable for algorithms with distinctive section divisions and explicit physical interpretations\cite{10.1145/1236463.1236468}. Unfortunately, the majority of signal processing algorithms do not always exhibit such clear-cut features. In contrast, the proposed AL-VPC divides an algorithm into basic arithmetic operations, making it applicable to any algorithm. In this context, conventional VPC emerges as a special case of AL-VPC.


Despite the considerable benefits, designing AL-VPC poses notable challenges, arising from the requirement of optimizing the arithmetic-level computing precision for each individual algorithm. In the following, we provide a detailed elaboration on the framework of AL-VPC.

\subsection{A New Storage Scheme: eBFP }

\begin{figure}
\centerline{\includegraphics[width=0.48\textwidth]{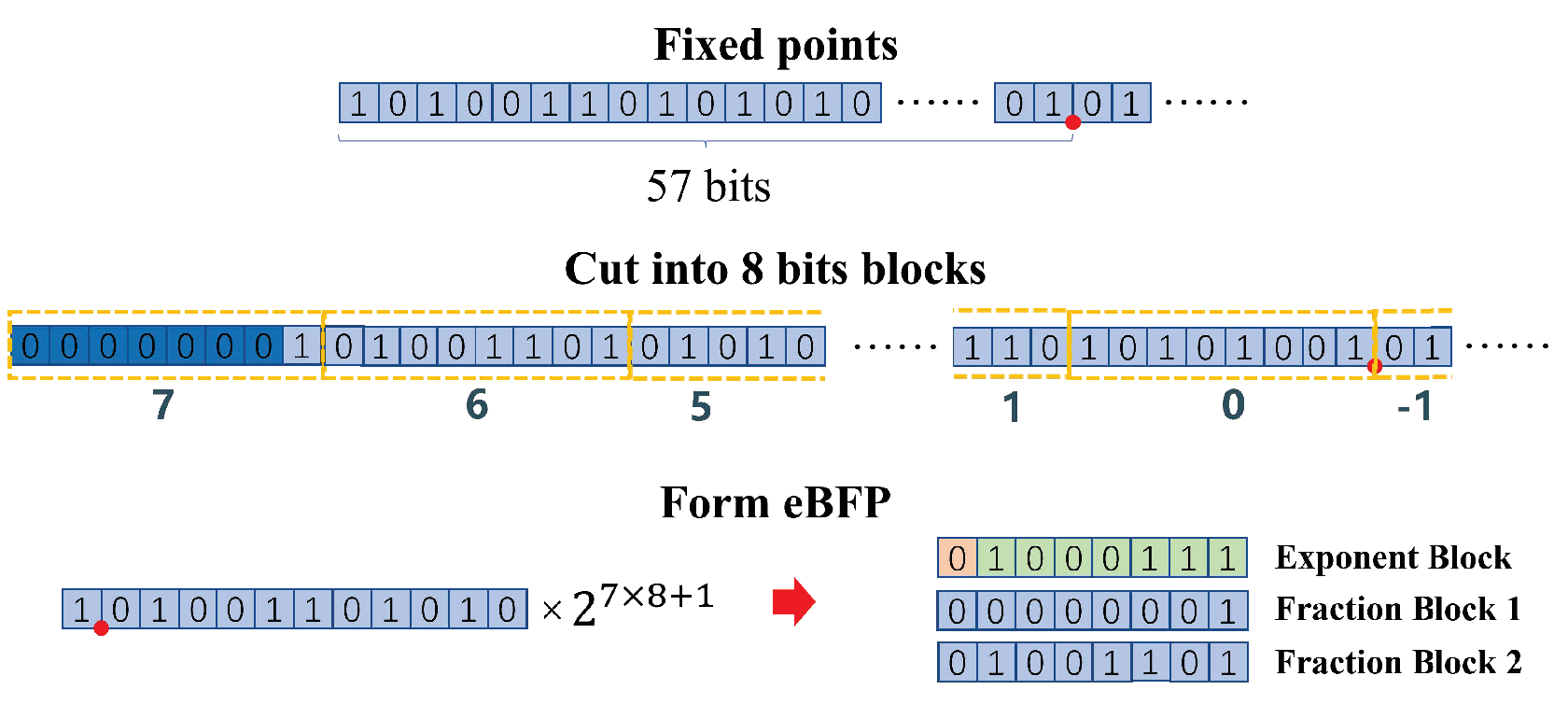}}
\caption{A demo of eBFP storage scheme, $F=8$, $E=8$, and $N=3$.}
\label{demoeBFP}
\end{figure}

In order to fully utilize the advantage of AL-VPC, a storage scheme with more precision options is favorable as it provides more freedom to optimize. However, existing storage scheme can not satisfy this demand, as conventional IEEE-754 standard only has four available precision, while fixed-point suffers from a critical constraint on the range of representable numbers. In addition, in-memory architectures inherently support only low bit-width computations, which are often insufficient to meet the accuracy requirements of signal processing algorithms. Therefore, we hereby design a brand new storage scheme, namely extended block floating point (eBFP), to support in-memory computing and AL-VPC.

eBFP divides the bits of a number into $F$ bits blocks, and define its exponent as the index of its first block. The first $F$ bits left to the decimal point is defined as the `zero' block with index 0. Other blocks are indexed from left to right with descending index number. Then, the index of the first block $e$ is then encoded as $e_{754}=e+2^{E-2}-1$ to become the exponent of the eBFP number, where $E$ is the bit width of the exponent. Finally,  first $N$ blocks are stored as fraction blocks with `0's added in front of the first block if needed. In total, eBFP has three parameters. $F$ controls the density of the available precision; $E$ controls the range of storage; while $N$ controls the precision of the number. We give a brief demonstration of the transition between a fixed-point number and an eBFP number in Fig.~\ref{demoeBFP}.

Compared to the standard IEEE-754, eBFP maintains a consistent width for exponent regardless of the number of blocks, which \textit{enables precision adjustments without altering the representation range}. This feature aligns well with practical needs, as precision control often necessitates maintaining a consistent representation range. The precision of eBFP is determined by the number
of blocks, offering finer control over precision adjustment compared to IEEE-754. Crucially, each eBFP block holds
a specific physical significance, facilitating storage and calculations at the block level. This feature enables in-memory architectures to achieve high-precision computation by combining multiple low-precision computations. 

We present a specification comparison between the configurations of eBFP and IEEE-754 in Table~\ref{ebfp}. For all eBFP specifications listed, $E=F=8$ is selected, and eBFP N$x$ denotes an eBFP number with $x$ fraction blocks. To employ eBFP into simulation, we also design each arithmetic operation from most basic logic operation, e.g., and, or, all the way to matrix operations, to enable MIMO signal processing algorithms. A more detailed introduction to the proposed eBFP is provided in Appendix~\ref{eBFP intro}.
\begin{table*}
\caption{Spec Comparison Between eBFP and IEEE-754 Standard}
\renewcommand\arraystretch{1.5}
\begin{center}
\begin{tabular}{|c|c|c|c|c|c|}
\hline
& Total bit width (bit) & Exponent  (bit)& Fraction (bit)& Maximum value (E) & Relative error\\ \hline
IEEE-754 half-precision & 16 & 5 & 10 & 4.51 & 9.77e-4 \\ \hline
eBFP N3 & 24 & 7 & 16 & 154.13 & 9.72e-4 \\ \hline
IEEE-754 single-precision & 32 & 8 & 23 & 38.23 & 1.19e-7 \\ \hline
eBFP N5 & 40 & 7 & 32 & 154.13 & 1.48e-8 \\ \hline
IEEE-754 double-precision & 64 & 11 & 52 & 307.95 & 2.22e-16 \\ \hline
eBFP N9 & 72 & 7 & 64 & 154.13 & 3.46e-18 \\ \hline
\end{tabular}\label{ebfp}
\end{center}
\end{table*}

\subsection{Expression Tree}

The second step in executing AL-VPC involves the segmentation of an algorithm into its basic arithmetic operations. Specifically, common linear precoding schemes in MIMO systems are composed of basic arithmetic operations, i.e., addition, subtraction, multiplication, division, and square root. Other arithmetic operations, e.g., complex and matrix arithmetics, can be broken down into a sequence of these basic arithmetic operations. In addition,  regardless of the initial input data, all algorithms inherently structure their arithmetic operations in a specific sequence. In computer science, the arithmetic operations of an algorithm are often represented using rooted trees \cite[Sec. 8.3]{10.5555/579402}, which are also known as expression trees. For any given algorithm, a corresponding expression tree can always be constructed.

\begin{figure*}
\centering
\includegraphics[width=1\textwidth]{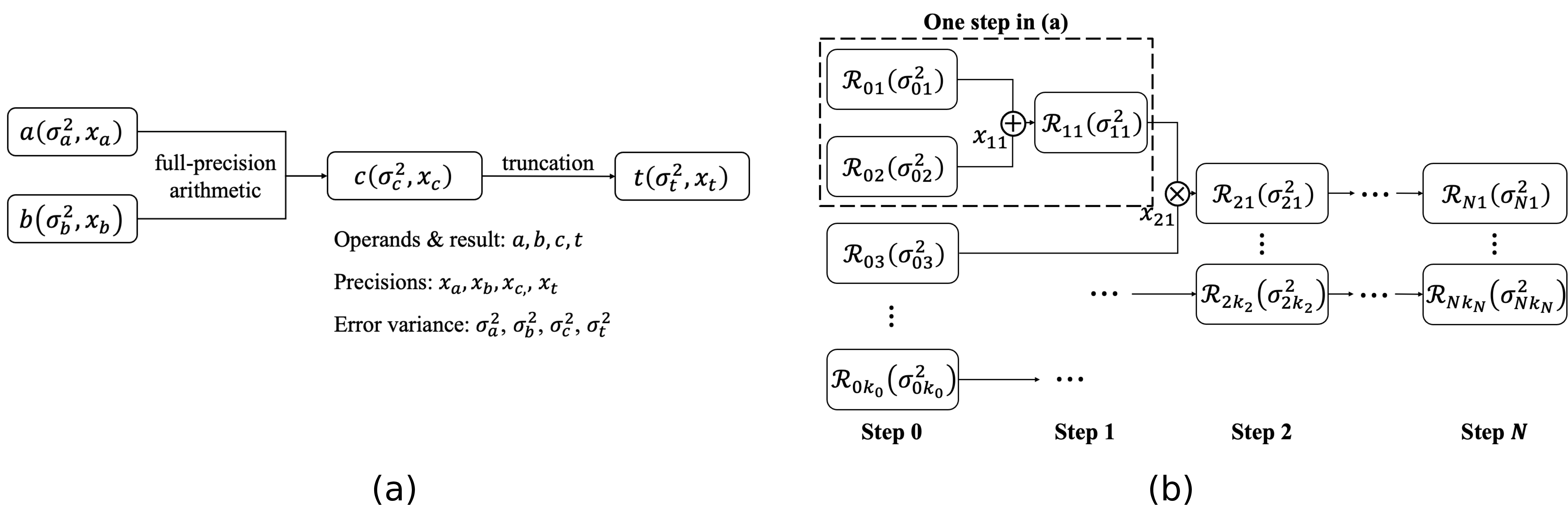}
\caption{(a) Error propagation during a one-step arithmetic; (b) Example of an expression tree of an $N$-step algorithm.}
\end{figure*}

Fig.~1 shows an example of an expression tree exploited in AL-VPC. In Fig.~1(a), the error propagation of a one-step finite-precision arithmetic is depicted with two operands. The arithmetic operation includes two steps. The first step is full-precision arithmetic which takes $a$ and $b$ as the inputs and outputs $c$, while the second step truncates $c$ to the desired precision and stores it as $t$. Denote the the precision of each number, i.e., bit width of the fraction of the number, by $x_a$, $x_b$, $x_c$, and $x_t$, respectively. Note that $x_c$ for the full-precision arithmetic is defined as the minimal computing precision that does not introduce new errors to the calculation. Further denote the corresponding relative errors of each number by $A$, $B$, $C$, and $T$ with variance of $\sigma_{a}^{2}$, $\sigma_{b}^{2}$, $\sigma_{c}^{2}$, and $\sigma_{t}^{2}$, respectively. During the process of this arithmetic operation, our primary focus on the variances of the errors. In the first step of full-precision arithmetic, $\sigma_{c}^{2}$ is related to $\sigma_{a}^{2}$ and $\sigma_{b}^{2}$, which can be treated as a transformation of random variables from $A$ and $B$ to $C$. In the second step of rounding, $\sigma_{t}^{2}$ depends on $\sigma_{c}^{2}$ and $x_t$.


Given the error propagation of the one-step arithmetic operation in Fig.~1(a), Fig.~1(b) shows the expression tree for a specific algorithm. In particular, $\sigma_{a}^{2}$, $\sigma_{b}^{2}$, and $\sigma_{t}^{2}$ in Fig.~1(a) correspond to $\sigma_{01}^{2}$, $\sigma_{02}^{2}$, and $\sigma_{11}^{2}$, respectively, in Fig.~1(b). We adopt ${\cal R}_{nk}$ to define the operand, where $n\in\{1,\dots,N\}$ refers to the number of steps and $k$ denotes the index of the operand in the step. Also, ${\cal R}_{0j},j\in\{1,2,3,\cdots,k_{0}\}$, are the $k_{0}$ initial operands given to the algorithm. $x_{nk}$ is defined as the computing precision of the respective arithmetic operation. $N$ is the number of sequential steps to obtain the final result. Note that initial inputs to the algorithm also consist error, as they are results of previous algorithms.



With the structure of the expression tree, an optimization problem can be formulated to address the trade-off between computing precision and computational complexity. In the following subsection, we start with modelling the one-step arithmetic operation in Fig.~1(a) and then proceed to derive an extension to multi-step arithmetic operations in Fig.~1(b).

\subsection{Optimization for One-step Arithmetic Operation}

To start, we define $o_{u}\left(x\right)$ as a measure of computational complexity, where $x$ represents computing precision of the arithmetic and $u\in\{\text{addition}, \text{subtraction},$ $ \text{multiplication}, \text{division}, \text{square root}\}$. Though $o_{u}$ differs among arithmetics, it is monotonically increasing with respect to (w.r.t.) the computing precision. While $x\in\mathbb N$, for the ease of tractability, we might temporarily treat it as continuous to facilitate analysis such that
\begin{equation}
\frac{\partial o_{u}}{\partial x_{}} \geq 0,\ \forall u
\end{equation}
holds in general. In addition, we also assume that $o_u$ is convex w.r.t. $x$.

Next, we can establish the relationship between $\sigma_{t}^{2}$ and $\{\sigma_{a}^{2}, \sigma_{b}^{2}, x\}$ as
\begin{equation}
\frac{\partial \sigma_{t}^{2}}{\partial \sigma_{a}^{2}} \geq 0,\ \frac{\partial \sigma_{t}^{2}}{\partial \sigma_{b}^{2}} \geq 0,\ \frac{\partial \sigma_{t}^{2}}{\partial x} \leq 0, 
\end{equation}
which hold because the error variance of the result is positively related to the error variance of operands and negatively related to the computing precision.

We then formulate a multi-objective optimization problem for the one-step arithmetic operation in Fig.~1(a) as
\begin{equation}
\begin{aligned}
\underset{x_{}}{\text{minimize}}\quad &\sigma_{t}^{2}\left(\sigma_{a},\sigma_{b},x\right), \\
\underset{x_{}}{\text{minimize}}\quad  &o_{u}\left(x\right) \\
\text {subject to} \quad &x_{\text {min}} \leq x_{} \leq x_{\text {max}},\
x_{} \in \mathbb{N},\label{3}
\end{aligned}
\end{equation}
where $x_{\text {min}}$ and $x_{\text {max}}$ are the minimum and maximum available computing precision of the computing device, which follows hardware constraint and subjective control.

In general, the Pareto boundary can be exploited to characterize the trade-off between the two objectives in \eqref{3}. Suppose there is a general utility function, $G\left(\sigma_{t}^{2},o_{u}\right)$, that serves as an equivalent single objective to replace the dual objectives. In practice, given that the form of $G$ does not affect the optimality, we define it as
\begin{equation}
    G=\sigma_{t}^2(x)+\alpha o_u(x),\label{3-1}
\end{equation}
where $\alpha\in(0,\infty)$ represents an indicator for balancing computing precision and complexity. With this explicit form of $G$, we have the convexity discussed in the following lemma.
\begin{lemma}\label{theorem of theorem 1}
$G$ of form \eqref{3-1} is convex w.r.t. $x$.
\end{lemma}

\noindent\textit{Proof:}
We start with the first-order derivative of $G$ as
\begin{equation}
\frac{\partial G}{\partial x} = \underbrace{\frac{\partial G}{\partial \sigma_{t}^2}\frac{\partial \sigma_{t}^2}{\partial x}}_{<0}+\underbrace{\frac{\partial G}{\partial o_{u}}\frac{\partial o_{u}}{\partial x}}_{>0},
\end{equation}
where $\frac{\partial G}{\partial \sigma_{t}^2}=1$, and $\frac{\partial G}{\partial o_{u}}>0$. 

Then, we have the second-order derivative of $G$ w.r.t. $x$ as
\begin{equation}
\begin{aligned}
\frac{\partial^{2} G}{\partial x^{2}}&=\frac{\partial G}{\partial \sigma_{t}^2}\frac{\partial^{2} \sigma_{t}^2}{\partial x^{2}}>0,\label{r15-1}
\end{aligned}
\end{equation}
where $\frac{\partial^{2} \sigma_{t}^2}{\partial x^{2}}>0$ comes from \eqref{100}. From \eqref{r15-1}, $G$ is a convex function.
$\hfill\square$


\subsection{Optimization for a Multi-step Algorithm}

For a complete algorithm, arithmetic operations are usually arranged in a specific order, from which an expression as depicted in Fig.~1(b) can be established. Analogously to \eqref{3}, we can formulate an optimization problem for computing precision assignment in multi-step algorithms as follows:
\begin{equation}
\begin{aligned}
\underset{x_{nk}}{\text{minimize}} \quad &F\left(\sigma_{N 0}^{2}, \cdots, \sigma_{N k_{N}}^{2}\right), \\
\underset{x_{nk}}{\text{minimize}} \quad &\sum_{n=1}^{N} \sum_{k=1}^{k_{n}} o_{u,nk} \\
\text {subject to } \quad &x_{nk,\text{min}} \leq x_{nk} \leq x_{nk,\text{max}}, \\
 &x_{nk} \in \mathbb{N},\ n=1, \cdots, N,\ k=1, \cdots, k_{n},\label{4}
\end{aligned}
\end{equation}
where $F\left(\cdot\right)$ captures the relationship between the performance of the algorithm and the error of the final result, $o_{u,nk}$ and $x_{nk}$ are the computational complexity and computing precision of the arithmetic with ${\cal R}_{nk}$, respectively.to  Similarly, we use an utility function $G_N$ to trade-off between two objectives
\begin{equation}
    G_N=F\left(\sigma_{N 0}, \cdots, \sigma_{N k_{N}}\right)+\alpha \sum_{n=1}^{N} \sum_{k=1}^{k_{n}} o_{u,nk}.
\end{equation}

With the general formulation established, we first analyze the arithmetic propagation error in Section \uppercase\expandafter{\romannumeral3} and then work on the solution to problem \eqref{4} in Section \uppercase\expandafter{\romannumeral4}.

\section{Arithmetic Propagation Error based on Stochastic Analysis}
The key to solving problem \eqref{3} and \eqref{4} is to explicitly obtain the expression of the variance of the error, i.e., the mathematical relationship between $\sigma_{t}^{2}$ and $\{\sigma_{a}^{2},\sigma_{b}^{2},x\}$ in \eqref{3}. In this section, we analyze the stochastic characteristic of one-step propagation error in Fig.~1(a). In the sequel, we first derive the error of full-precision arithmetics by performing the transformation of the random variables to obtain $\sigma_{c}^2$, as well as the rounding error to obtain $\sigma_{t}^{2}$. In the following, we assume that all the variance of the relative errors is much smaller than 1, which is well-justified in practice.

\subsection{Error of Full-Precision Arithmetic}
We directly present the error of full-precision arithmetic in Lemma \ref{theorem of full precision arithmetic} by deriving the essential statistics of $\sigma_{c}^2$.

\begin{lemma}\label{theorem of full precision arithmetic}
The mean and variance of the error of full-precision arithmetic are listed in Table~\uppercase\expandafter{\romannumeral 1}.

\begin{table}
\caption{$\mathbb E\{C\}$ and  $\sigma_{c}^{2}$ of Basic Arithmetics}
\renewcommand\arraystretch{1.5}
\begin{center}
\begin{tabular}{ccc}
\toprule
Arithmetic & Mean & Variance\\
\midrule
Addition & 0 & $\left(a^{2}\sigma_{a}^{2}+b^{2}\sigma_{b}^{2}\right)/\left(a+b\right)^{2}$\\
Subtraction & 0 & $\left(a^{2}\sigma_{a}^{2}+b^{2}\sigma_{b}^{2}\right)/\left(a-b\right)^{2}$\\
Multiplication & 0 & $\sigma_{a}^{2}+\sigma_{b}^{2}\left(+\sigma_{a}^{2}\sigma_{b}^{2}\right)$\\
Division & 0 & $\sigma_{a}^{2}+\sigma_{b}^{2}$\\
Square root & 0 & $\sigma_{a}^{2}/4$\\
\bottomrule
\end{tabular}\label{tab1}
\end{center}
\end{table}
\end{lemma}

\noindent\textit{Proof:}
We first derive the variance of $C$. Since the proof for subtraction is very similar to the one of addition, it is omitted for brevity.

\subsubsection{Addition}

For an addition of two operands, say, $a+b$, the resulting relative error is expressed as

\begin{equation}
    C=\frac{a\left(1+A\right)+b\left(1+B\right)-\left(a+b\right)}{a+b}=\frac{aA+bB}{a+b},
\end{equation}
and the variance of $C$ is

\begin{equation}
    \sigma_{c}^{2}=\mathbb{V}\left[\frac{aA+bB}{a+b}\right]=\frac{a^{2}\sigma_{a}^{2}+b^{2}\sigma_{b}^{2}}{\left(a+b\right)^{2}},
\end{equation}
where the definitions of variables refer to Section~\uppercase\expandafter{\romannumeral2}-A, and $\mathbb{V}\left[\cdot\right]$ represents the variance operator. 

\subsubsection{Multiplication}

For a multiplication of two operands, say, $ab$, it follows

\begin{equation}
C=\frac{a\left(1+A\right)b\left(1+B\right)-ab}{ab}=A+B+AB.
\end{equation}
Then, the variance of $C$ equals
\begin{align}
\sigma_{c}^{2}&=\left(1+2\mathbb{E}\left[B\right]\right)\sigma_{a}^{2}+\left(1+2\mathbb{E}\left[A\right]\right)\sigma_{b}^2\notag\\
&\quad+\sigma_{a}^{2}\sigma_{b}^{2}+\left(\mathbb{E}\left[A\right]\right)^{2}\sigma_{b}^{2}+\left(\mathbb{E}\left[B\right]\right)^{2}\sigma_{a}^{2}.
\end{align}

 Supposing that $\mathbb{E}\left[A\right]=\mathbb{E}\left[B\right]=0$ (this assumption will be discussed later), we have
 \begin{equation}
 \sigma_{c}^{2}=\sigma_{a}^{2}+\sigma_{b}^{2}+\sigma_{a}^{2}\sigma_{b}^{2}.
 \end{equation}
Note that if the computing precision is relatively high, $\sigma_{a}^{2}\sigma_{b}^{2}$ is negligible.
\subsubsection{Division}

For a division of two operands, say, $a/b$, it follows

\begin{equation}
C=\frac{\frac{a\left(1+A\right)}{b\left(1+B\right)}-\frac{a}{b}}{\frac{a}{b}}=\frac{A-B}{1+B},
\end{equation}
and the variance of $C$ is
\begin{align}
\sigma_{c}^{2}&=\mathbb{V}\left[\frac{A}{1+B}\right]+\mathbb{V}\left[\frac{B}{1+B}\right]\notag\\
&\quad-2\operatorname{Cov}\left(\frac{A}{1+B},\frac{B}{1+B}\right),\label{203}
\end{align}
where $\operatorname{Cov}\left(\cdot,\cdot \right)$ represents the covariance operator.

We then derive $\mathbb{V}\left[\frac{X}{Y}\right]$. Applying the first-order Taylor expansion to $f(X, Y)=X / Y$ at  $ \delta=(\mathbb{E}\left[X\right], \mathbb{E}\left[Y\right])$, we obtain 
\begin{align}
\frac{X}{Y}&=f(X, Y)=f(\delta)+f_{X}^{\prime}(\delta)(X-\mathbb{E}\left[X\right])\notag\\
&\quad+f_{Y}^{\prime}(\delta)(Y-\mathbb{E}\left[Y\right])+O\left(X,Y\right),\label{50}    
\end{align}
where $O\left(X,Y\right)$ is a remainder of smaller order. Then, the expectation and variance can be obtained as
\begin{align}
\mathbb{E}\left[\frac{X}{Y}\right]&\approx f(\delta)+f_{X}^{\prime}(\delta) \mathbb{E}\left[X-\mathbb{E}\left[X\right]\right]+f_{Y}^{\prime}(\delta) \mathbb{E}\left[Y-\mathbb{E}\left[Y\right]\right]\notag\\
&=\frac{\mathbb{E}\left[X\right]}{\mathbb{E}\left[Y\right]},\label{51}   
\end{align}

\begin{equation}
\begin{aligned}
\mathbb{V}\left[\frac{X}{Y}\right]&=\mathbb{E}\left[\left[f_{X}^{\prime}(\delta)(X-\mathbb{E}\left[X\right])+f_{Y}^{\prime}(\delta)(Y-\mathbb{E}\left[Y\right])\right]^{2}\right]\\
&=\left(\frac{\mathbb{E}\left[X\right]}{\mathbb{E}\left[Y\right]}\right)^{2}\bigg[\frac{\mathbb{V}\left[X\right]}{\left(\mathbb{E}\left[X\right]\right)^{2}}+\frac{\mathbb{V}\left[Y\right]}{\left(\mathbb{E}\left[Y\right]\right)^{2}}\\
&\quad-2 \frac{\operatorname{Cov}\left(X, Y\right)}{\mathbb{E}\left[X\right] \mathbb{E}\left[Y\right]}\bigg].\label{52}
\end{aligned}
\end{equation}

Therefore, we have
\begin{align}
\mathbb{V}\left[\frac{A}{1+B}\right]&=\frac{\sigma_{a}^{2}}{(1+\mathbb{E}\left[B\right])^{2}}+\frac{(\mathbb{E}\left[A\right])^{2} \sigma_{b}^{2}}{(1+\mathbb{E}\left[B\right])^{4}}\notag\\
&=\sigma_{a}^{2},\label{201}
\end{align}
and similarly $\mathbb{V}\left[\frac{B}{1+B}\right]=\sigma_{b}^{2}$. In addition, assuming $\mathbb{E}\left[A\right]=0$, it follows

\begin{equation}
\operatorname{Cov}\left(\frac{A}{1+B}, \frac{B}{1+B}\right)=\mathbb{E}\left[A\right] \operatorname{Cov}\left(\frac{1}{1+B}, \frac{B}{1+B}\right)=0.\label{202}
\end{equation}
Now, by substituting \eqref{201} and \eqref{202} into \eqref{203}, it yields
\begin{equation}
\sigma_{c}^{2}=\sigma_{a}^{2}+\sigma_{b}^{2}.
\end{equation}
\subsubsection{Square root }

For a square root of one operand, say, $\sqrt{a}$, it follows
\begin{equation}
C=\frac{\sqrt{a\left(1+A\right)}-\sqrt{a}}{\sqrt{a}}=\sqrt{1+A}-1.
\end{equation}
With the techniques similar to division, the variance of $C$ is given as
\begin{align}
\sigma_{c}^{2}&=\mathbb{V}\left[\sqrt{1+A}\right]=\frac{\mathbb{V}\left[1+A\right]}{\left(2\sqrt{\mathbb{E}\left[1+A\right]}\right)^2}=\frac{1}{4}\sigma_{a}^{2}.
\end{align}

Now we derive the mean of $C$. Assuming $\mathbb{E}\left[A\right]=\mathbb{E}\left[B\right]=0$, it is easy to check that $\mathbb{E}\left[C\right]=0$ for all arithmetics. It implies that if the error of initial input numbers have zero means, all the results have statistically zero-mean errors. 
$\hfill\square$

\begin{remark}
Variances of addition and subtraction depend on the values of their operands, while the variances of multiplication, division, and square root do not rely on their operands. 
\end{remark}

The relative error variances for multiplication, division, and square root are independent of the operand
values. These operations exhibit a more predictable error behavior under relative metrics, enabling their error
variances to be estimated or bounded analytically, even before the actual inputs are known. This important remark will be exploited in Section~\uppercase\expandafter{\romannumeral 4} to facilitate the solution to problem \eqref{4}.

\subsection{Error of Rounding}

Given the error of the full-precision arithmetic, i.e., $\sigma_{c}^{2}$, in the previous subsection, we next derive the error introduced by rounding, i.e., the relationship between $\sigma_{c}^2$ and $\sigma_{t}^2$ in Fig~1(a). 





To characterize the distribution of the error introduced by rounding, we introduce a real-valued random variable $X$ and the corresponding rounding error $E$. It follows

\begin{equation}
    E=\frac{X-\mathrm{Round}\left(X\right)}{X},
\end{equation}
where
\begin{equation}
    \mathrm{Round}\left(n\right)=\{m\big||m-n|\le|m'-n|,\,\forall m'\in \mathbb F\},
\end{equation}
and $\mathbb{F}$ is the set of all representable numbers in the floating-point.

Since $E$ is a random variable, it is also common to represent $E$ as $E=r\left(x\right)W$, where $r\left(x\right)=\varepsilon^{-x-1}$ is the maximum rounding error, $W\in\left[-1,1\right]$ is a corresponding random variable \cite{9048893}, and $\varepsilon$ is the base of storage, e.g., $\varepsilon=2$ for binary storage. In this context, the statistical characteristics of $W$ is equivalent to those of $E$, which can be exploited to facilitate the following derivation.  Follow the definition of rounding error, we derive the relationship between $\sigma_{c}^2$ and $\sigma_{t}^2$ in the following theorem.

\begin{theorem}\label{theorem of rounding error}
For any arithmetic operation, the variance of rounding error is
\begin{align}
\sigma_{t}^{2} &=\left(1+r\left(x\right)^{2} \sigma_{W}^{2}+2r\left(x\right)\mathbb{E}\left[W\right]+r\left(x\right)^{2} \mathbb{E}\left[W\right]^{2}\right) \sigma_{c}^{2}\notag\\
&\quad+r\left(x\right)^{2} \sigma_{W}^{2},    \label{100}
\end{align}
where $\mathbb{E}\left[W\right]$ and $\sigma_{W}^{2}$ are respectively the expectation and variance of $W$.  When the computing precision is sufficiently high, we have a tight approximation as

\begin{equation}
\sigma_{t}^{2} \approx \left(1+r\left(x\right)^{2} \sigma_{W}^{2}\right) \sigma_{c}^{2}+r\left(x\right)^{2} \sigma_{W}^{2}. \label{12321}
\end{equation}

\end{theorem}

\noindent\textit{Proof:}
Following the definitions in Section \uppercase\expandafter{\romannumeral 2}-A and Section \uppercase\expandafter{\romannumeral 3}-A, the variance of the rounding error is expressed as
\begin{align}
\sigma_{t}^{2}&=\mathbb{V}\left[\frac{c\left(1+C\right)\left(1+r\left(x\right)W\right)-c}{c}\right]\notag\\
&=\left(1+r\left(x\right)^{2} \sigma_{W}^{2}+2r\left(x\right)\mathbb{E}\left[W\right]+r\left(x\right)^{2} \mathbb{E}\left[W\right]^{2}\right) \sigma_{c}^{2}\notag\\
&\quad+r\left(x\right)^{2} \sigma_{W}^{2}.
\end{align}

When $x\rightarrow\infty$, it has been proved in \cite{9048893} that the PDF of $W$ is \eqref{0}.

\begin{figure*}
\begin{equation}
\lim _{x \rightarrow \infty} f_{W}(w)=\left\{\begin{array}{cl}
\frac{3}{4}, &w \in\left[-\frac{1}{2}, \frac{1}{2}\right) \\
\frac{1}{2}\left(\frac{1}{w}-1\right)+\frac{1}{4}\left(\frac{1}{w}-1\right)^{2}, &w \in\left[-\frac{1}{2}, \frac{1}{2}\right)\cup\left(\frac{1}{2}, 1\right]\label{0}
\end{array}\right.
\end{equation}
\hrulefill
\end{figure*}

Note that when $x$ is sufficiently large, $f_{W}\left(w\right)$ converges and becomes independent of $x$, so does $\sigma_{W}^{2}$, which means the rounding error becomes independent with the signal. Table~\ref{table4} shows the convergence values of $\sigma_{W}^{2}$  and $\mathbb{E}\left[W\right]$, when $\varepsilon$ is set to 2 for the binary case.

\begin{table}
\caption{$\sigma_{W}^{2}$ and $\mathbb{E}\left[W\right]$ with respect to $x$ ($\varepsilon=2$)}
\label{table4}
\begin{center}
\begin{tabular}{cccccccc}
\toprule
$x$ & 1 & 3 & 5 & 7 & 9 & 11 & 13\\
\midrule
$\sigma_{W}^{2}$ & 0.353 & 0.213 & 0.179 & 0.170 & 0.167 & 0.167 & 0.167\\[1.5ex]
$\mathbb{E}\left[W\right]$ & -0.09 & -6e-3 & -4e-4 & -3e-5 & -2e-6 & -1e-7 & -6e-9\\[1.5ex]
\bottomrule
\end{tabular}
\end{center}
\end{table}

It is obvious that in the binary case, $\sigma_{W}^{2}$ converges to 0.167 and $\mathbb{E}\left[W\right]$ converges to zero. For other bases, similar results can be obtained. Consequently, the approximation in \eqref{12321} is obtained.
$\hfill\square$

\begin{remark}
The calculation of the variance of the rounding error can be separated from the calculation of the variance of the full-precision arithmetic.
\end{remark}
This remark confirms that the propagation error of a one-step arithmetic operation can be derived with two distinct steps.

Thus far, we have derived a stochastic model for characterizing the one-step arithmetic propagation error as illustrated in Fig.~1(a), which is exploited in the explicit formulation of problem \eqref{4} of Fig.~1(b).

\section{Problem Formulation and Optimization for Multi-step Algorithm}

In this section, we develop two algorithms to address the optimization problem of computing precision assignment for a multi-step algorithm as described in problem \eqref{4}. Unlike the problem of the one-step arithmetic in \eqref{3}, the convexity of problem \eqref{4} is generally unclear, which means that it is generally challenging to solve with conventional methods.

\subsection{Explication of Problem \eqref{4}}

\begin{figure*}
\begin{equation}
\frac{\partial G_{N}}{\partial x_{nk}}=\frac{\partial G_{N}}{\partial F}\frac{\partial F}{\partial \sigma_{t,Nk_{N}}}\left(\prod_{i=0}^{N-n-1}\underbrace{\frac{\partial \sigma_{t,\left(N-i\right)k_{N-i}}^{2}}{\partial \sigma_{c,\left(N-i\right) k_{N-i}}^{2}}}_{\circled{1}}\underbrace{\frac{\partial \sigma_{c,\left(N-i\right) k_{N-i}}^{2}}{\partial \sigma_{t,\left(N-i-1\right)k_{N-i-1}}^{2}}}_{\circled{2}}\right)\underbrace{\frac{\partial \sigma_{t,nk}^{2}}{\partial x_{nk}}}_{\circled{3}}+\frac{\partial G_{N}}{\partial o_{u}}\frac{\partial o_{u}}{\partial x_{nk}}\label{5}
\end{equation}
\hrulefill
\end{figure*}

\begin{figure*}
\begin{equation}
\frac{\partial \sigma_{41}^{2}}{\partial \sigma_{c,41}^{2}}\cdot\frac{\partial \sigma_{c,41}^{2}}{\partial \sigma_{31}^{2}}\cdot\frac{\partial \sigma_{31}^{2}}{\partial \sigma_{c,31}^{2}}\cdot\frac{\partial \sigma_{c,31}^{2}}{\partial \sigma_{22}^{2}}=\left(1+r\left(x_{41}\right)^{2}\sigma_{W}^{2}\right)\cdot\frac{{\cal R}_{31}^{2}}{\left({\cal R}_{31}+{\cal R}_{21}\right)^{2}}\cdot\left(1+r\left(x_{31}\right)^{2}\sigma_{W}^{2}\right)\cdot1\label{101}
\end{equation}
\hrulefill
\end{figure*}

To start with, we temporarily set aside the constraint of problem \eqref{4}. A sub-optimal solution can be obtained by examining $\partial G_{N}/\partial x_{nk}=0$. By applying the chain rule of derivatives, $\partial G_{N}/\partial x_{nk}$ is written in \eqref{5} following the  \textit{back propagation} of the arithmetic operations after ${\cal R}_{nk}$ in Fig.~1(b) with the circled values calculated in \eqref{7}, where $a$ and $b$ refer to the two operands of the arithmetic. Note that $\circled1$ is derived from \eqref{100}, $\circled2$ is obtained from Table~\uppercase\expandafter{\romannumeral1}, and $\circled3$ adopts \eqref{100} with $r\left(x\right)=\varepsilon^{-x-1}$.

\begin{align}
\circled{1}&=1+r x_{\left(N-i\right)k_{N-i}}^{2}\sigma_{W}^{2}\label{6}\\
\circled{2}&=\left\{\begin{array}{cl}
\frac{a^{2}}{\left(a+b\right)^{2}},&\text{addition}\\
\frac{a^{2}}{\left(a-b\right)^{2}},&\text{subtraction} \\
 ,\text{multiplication \& division}\\
\frac{1}{4},&\text{square root}
\end{array}\right.\label{6.5}\\
\circled{3}&=-2\varepsilon^{-2}\sigma_{W}^{2}\left(1+\sigma_{c,nk}^{2}\right)\varepsilon^{-2x_{nk}}\text{ln}\varepsilon,\label{7}
\end{align}

\begin{figure}
\centering
\includegraphics[width=0.48\textwidth,height=0.22\textwidth]{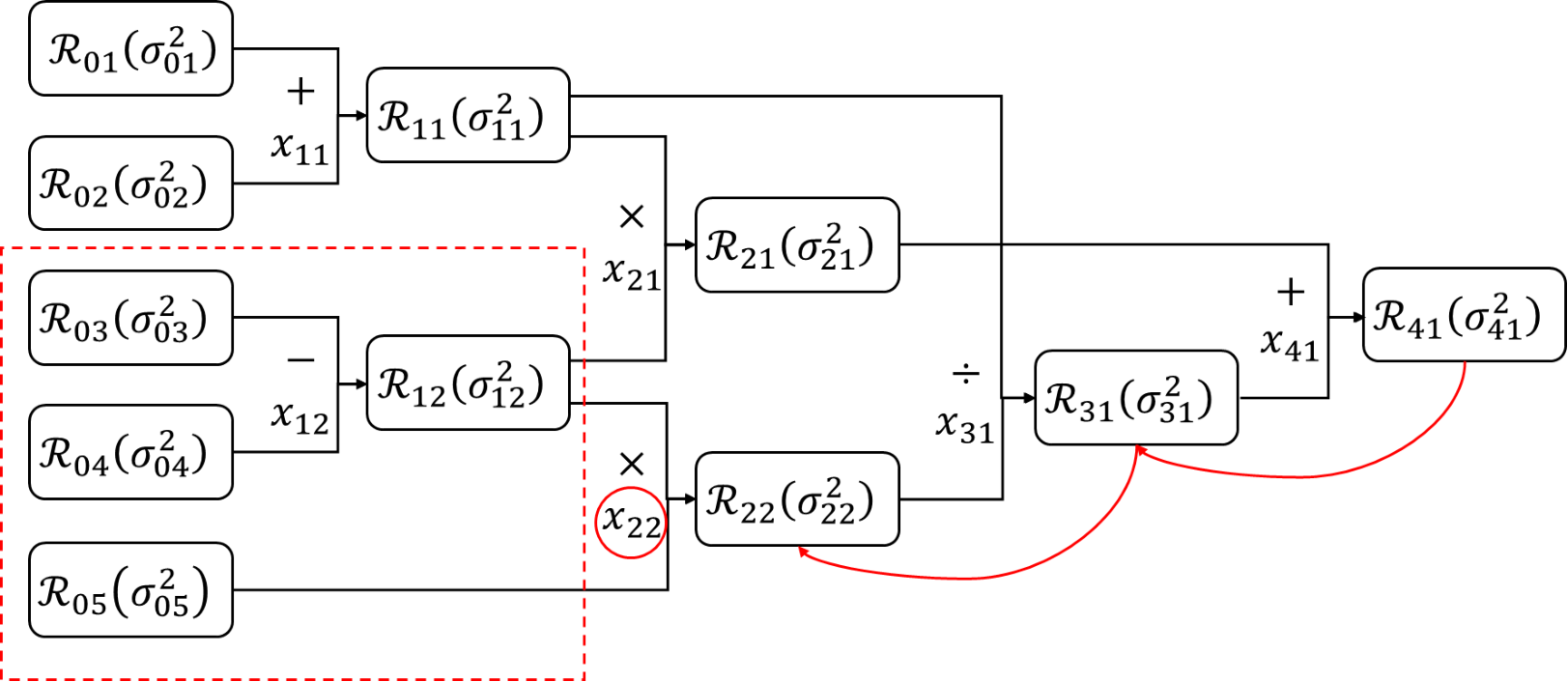}
\caption{An example of the solution to \eqref{5}.}
\label{fig2}
\end{figure}

The expression in \eqref{5} is a bit involved. To elucidate, let us consider an example from Fig.~\ref{fig2} which contains six arithmetic operations and aim to optimize $x_{22}$. We first compose $\partial G_{N}/\partial x_{22}=0$  using the back propagation from the final result of the algorithm to ${\cal R}_{22}$, which traces the path ${\cal R}_{41}\rightarrow {\cal R}_{31}\rightarrow {\cal R}_{22}$. With this process, the parts of the products in the bracket of \eqref{5} is calculated explicitly in \eqref{101}. While $\sigma_{c,nk}^{2}$  in \circled{3}, i.e., $\sigma_{c,22}^{2}$ is derived as

\begin{align}
\sigma_{c,22}^{2}&=\sigma_{12}^{2}+\sigma_{05}^{2}\notag\\
&=\sigma_{c,12}^{2}\left[1+r\left(x_{12}\right)^{2}\sigma_{W}^{2}\right]+r\left(x_{12}\right)^{2}\sigma_{W}^{2}+\sigma_{05}^{2}\notag\\
&=\frac{{\cal R}_{03}^{2}\sigma_{03}^{2}+{\cal R}_{04}^{2}\sigma_{04}^{2}}{\left({\cal R}_{03}-{\cal R}_{04}\right)^{2}}\left[1+r\left(x_{12}\right)^{2}\sigma_{W}^{2}\right]\notag\\
&\quad+r\left(x_{12}\right)^{2}\sigma_{W}^{2}+\sigma_{05}^{2}.\label{102}
\end{align}

The remaining  of \eqref{5} can be obtained similarly, so a complete expansion of \eqref{5} is not detailed for the sake of brevity.

We are now ready to calculate the optimized computing precision of the previous arithmetic operations. Extract parts without $x_{nk}$ and rewrite ${\partial G_{N}}/{\partial x_{nk}}\triangleq G_{\sigma,nk}\varepsilon^{-2x_{nk}}+G_{o}o_{u}^{(-1)}$, where $G_{\sigma,nk}$ and $G_{o}$ do not contain the term $x_{nk}$. The calculations of the optimized computing precision using back propagation is presented in Theorem~\ref{theorem of online vpc}.

\begin{theorem}\label{theorem of online vpc}
For any ${\cal R}_{nk}$, if ${G_{\sigma,nk}}$, ${G_{o}}$ and $x_{nk}^{\rm{opt}}$ are acknowledged, then for the next arithmetic ${\cal R}_{(n+1)k^{\prime}}$, we have

\begin{align}
\frac{G_{\sigma,(n+1) k^{\prime}}}{G_{o}}&=\frac{G_{\sigma, n k}}{G_{o}}\frac{\partial \sigma_{t,n k}^{2}}{\partial \sigma_{c,(n+1) k^{\prime}}^{2}},\label{8-1}\\
x_{(n+1) k^{\prime}}^{\mathrm{opt}}&=\mathrm{LUT}_{x^{\mathrm{opt}}}\left(G_{\sigma,(n+1) k^{\prime}} / G_{o}, u\right),\label{9}
\end{align}
where $\mathrm{LUT}_{x^{\mathrm{opt}}}\left(G_{\sigma,(n+1) k^{\prime}} / G_{o}, u\right)$ is a look up table (LUT) which returns $x_{(n+1) k^{\prime}}^{\mathrm{opt}}$ given $G_{\sigma,(n+1) k^{\prime}} / G_{o}$ and the type of arithmetic $u$.
\end{theorem}

\noindent\textit{Proof:}
By the definition given in Theorem 2, for $\mathcal R_{(n+1)k^{\prime}}$ we have
\begin{equation}
\frac{\partial G_{N}}{\partial x_{\left(n+1\right)k'}}=G_{\sigma,\left(n+1\right)k'}\varepsilon^{-2x_{\left(n+1\right)k'}}+G_{o}o_{u}^{(-1)}\left(x_{\left(n+1\right)k'}\right).\label{-202}
\end{equation}
Comparing $G_{\sigma,nk}$ and $G_{\sigma,\left(n+1\right)k'}$, it follows

\begin{align}
\frac{G_{\sigma,(n+1) k^{\prime}}}{G_{\sigma, nk}}&=\left(\frac{\partial \sigma_{t,(n+1) k^{\prime}}^{2}}{\partial \sigma_{c,(n+1) k^{\prime}}^{2}} \frac{\partial \sigma_{c,(n+1) m^{\prime}}^{2}}{\partial \sigma_{t,n k}^{2}}\right)^{-1} \frac{1+\sigma_{c,(n+1) k^{\prime}}^{2}}{1+\sigma_{c, n k}^{2}}\notag \\
&\overset{\left(\text{a}\right)}{\approx}\left[\left(1+r\left(x_{\left(n+1\right)k^{\prime}}\right)^{2}\sigma_{W}^{2}\right)\frac{\partial \sigma_{c,(n+1) k^{\prime}}^{2}}{\partial \sigma_{t,n k}^{2}}\right]^{-1}\notag\\
&\overset{\left(\text{b}\right)}{\approx}\frac{\partial \sigma_{t,n k}^{2}}{\partial \sigma_{c,(n+1) k^{\prime}}^{2}}.\label{-201}
\end{align}
where $\left(\text{a}\right)$ follows that both $\sigma_{c,\left(n+1\right)k^{\prime}}^{2}$ and $\sigma_{c,nk}^{2}$ are much smaller than 1, and $\left(\text{b}\right)$ follows that $r\left(x_{\left(n+1\right)k^{\prime}}\right)^{2}\sigma_{W}^{2}$ is much smaller than 1. Then, from \eqref{-201}, \eqref{8-1} is derived.

Meanwhile, to obtain the optimal $x_{nk}$, let $\partial G_{N}/\partial x_{nk}=0$ and denote
\begin{equation}
f_{u}\left(x_{nk}\right)\triangleq \frac{o_{u}^{(-1)}\left(x_{nk}\right)}{\varepsilon^{-2x_{nk}}}=-\frac{G_{\sigma,nk}}{G_{o}}.
\end{equation}
Since $x_{nk}$ is discrete and $f_{u}\left(x_{nk}\right)$ is strictly monotonically increasing, denote the optimal $x_{nk}$ by $x_{nk}^{\text{opt}}$, a look-up table (LUT) between ${G_{\sigma,nk}}/{G_{o}}$, $u$ and $x_{nk}^{\text{opt}}$ can be established, denoted by $x_{nk}^{\text{opt}}=\text{LUT}_{x^{\text{opt}}}\left({G_{\sigma,nk}}/{G_{o}},u\right)$. Then, the LUT for $x_{(n+1) k^{\prime}}^{\mathrm{opt}}$ in \eqref{9} is obtained
$\hfill\square$

Since ${\partial \sigma_{t,n k}^{2}}/{\partial \sigma_{c,(n+1) k^{\prime}}^{2}}$ depends on the operands of the arithmetic operation, which are unknown to the back propagation as the arithmetic operations have not yet been calculated, we approximate ${\partial \sigma_{t,n k}^{2}}/{\partial \sigma_{c,(n+1) k^{\prime}}^{2}}$ by its expectation $\mathbb{E}\left\{{\partial \sigma_{t,n k}^{2}}/{\partial \sigma_{c,(n+1) k^{\prime}}^{2}}\right\}$. In this manner, with the optimal computing precision of the last arithmetic operations, the optimized computing precision for all the arithmetic operations can be calculated through back propagation according to Theorem \ref{theorem of online vpc}. In this method, the optimized computing precision for all the arithmetic operations is obtained before the execution of the algorithm. We refer to this method as \textit{Offline VPC}. 

On the other hand, we can alternatively perform a forward propagation instead of backward propagation, obtaining the optimized computing precision of each arithmetic operation during its calculation. In this way, the operands for each arithmetic operation are precisely known. Theoretically, this approach can achieve a better trade-off result than offline VPC. Since the optimal computing precision is calculated while the execution of the algorithm, we refer to this method as \textit{Online VPC}.  In the following, we develop the procedures and algorithms for offline VPC and online VPC, respectively.

\subsection{Offline VPC}

Since for multiplication, division, and square root, ${\partial \sigma_{t,n k}^{2}}/{\partial \sigma_{c,(n+1) k^{\prime}}^{2}}$ does not depend on its operands, we only need to calculate the expectation of ${\partial \sigma_{t,n k}^{2}}/{\partial \sigma_{c,(n+1) k^{\prime}}^{2}}$ for addition and subtraction that leads to the following lemma.

\begin{lemma}\label{theorem of speculation}
The average error of the full-precision arithmetic of a one-step addition or subtraction is
\begin{align}
\mathbb{E}&\left[\frac{\partial \sigma_{t,n k}^{2}}{\partial \sigma_{c,(n+1) k^{\prime}}^{2}}\right]\notag\\
&\approx\left\{\begin{array}{cl}
1+\frac{1}{\ln 2} 2^{2-e_{b}}>1,&\text{addition} \\
1 - \frac{1}{2^{e_{b}-2} \ln 2-\frac{\ln 2}{2}-\frac{1}{4}}<1,&\text{subtraction}, \\
\end{array}\right.\label{7.5}
\end{align}
where $e_{b}$ is the bit width of the exponent of the floating point.
\end{lemma}

\noindent\textit{Proof:}
First, we consider the expectation of ${\partial \sigma_{t,n k}^{2}}/{\partial \sigma_{c,(n+1) k^{\prime}}^{2}}$ for addition. Since the addition between positive operand $x$ and negative operand $y$ is equivalent to the subtraction between two positive operands $x$ and $-y$, without loss of generality, we only consider the addition with two positive operands.

Suppose that the two operands are $X$ and $Y$, which both uniformly distribute in $\left[{\cal R}_{\text{min}},{\cal R}_{\text{max}}\right]$, where  ${\cal R}_{\text{min}}$ and ${\cal R}_{\text{max}}$ are the smallest and largest positive numbers that can be stored. Denote $Z=Y/X$, then

\begin{equation}
\mathbb{E}\left[\frac{\partial \sigma_{c_{t},n k}^{2}}{\partial \sigma_{c,(n+1) k^{\prime}}^{2}}\right]=\mathbb{E}\left[\frac{(Z+1)^{2}}{Z^{2}}\right]=1+\frac{2}{\mathbb{E}[Z]}+\frac{1}{\mathbb{E}\left[Z^{2}\right]}.\label{-205}
\end{equation}

Even though $Z$ is discrete in theory, it can be treated as continuous because the density is very fine. For $Z=Y/X$, the PDF of $Z$ is

\begin{equation}
f_{Z}(z)=\int_{D}|x| f_{X}(x) f_{Y}(x z)\ \text{d} x.\label{-2}
\end{equation}
The region of the integral in \eqref{-2} is

\begin{equation}
D= \left\{x\bigg| x\in\left[{\cal R}_{\text{min}},{\cal R}_{\text{max}}\right] \ \text{and} \ xz\in\left[{\cal R}_{\text{min}},{\cal R}_{\text{max}}\right]\right\}.
\end{equation}
According to the range of $z$, the range of $x$ is

\begin{equation}
x\in\left\{\begin{array}{cl}
\left[{\cal R}_{\text{min}}, \frac{{\cal R}_{\text{max}}}{z}\right],&1\le z\leq \frac{{\cal R}_{\text{max}}}{{\cal R}_{\text{min}}} \\
\left[\frac{{\cal R}_{\text{min}}}{z},{\cal R}_{\text{max}}\right],&\frac{{\cal R}_{\text{min}}}{{\cal R}_{\text{max}}}\le z<1.
\end{array}\right.
\end{equation}
Then, when $1\le z\leq \frac{{\cal R}_{\text{max}}}{{\cal R}_{\text{min}}}$, we have

\begin{equation}
\begin{aligned}
f_{Z}\left(z\right)=\frac{\frac{{\cal R}_{\max }^{2}}{z^{2}}-{\cal R}_{\min }^{2}}{2\left({\cal R}_{\max }-{\cal R}_{\min }\right)^{2}},
\end{aligned}
\end{equation}
the remaining part can be obtained similarly. To conclude
\begin{equation}
f_{Z}(z)=\left\{\begin{array}{cl}
\frac{{\cal R}_{\max }^{2}-\frac{{\cal R}_{\min }^{2}}{z^{2}}}{2\left({\cal R}_{\max }-{\cal R}_{\min }\right)^{2}}, &\frac{{\cal R}_{\min }}{{\cal R}_{\max }} \leq z<1 \\
\frac{\frac{{\cal R}_{\max }^{2}}{z^{2}}-{\cal R}_{\min }^{2}}{2\left({\cal R}_{\max }-{\cal R}_{\min }\right)^{2}}, &1 \leq z \leq \frac{{\cal R}_{\max }}{{\cal R}_{\min }}.\label{10}
\end{array}\right.
\end{equation}
From the definition of floating points\cite{4610935}, ${\cal R}_{\max}/{\cal R}_{\min}=2^{2^{e_{b}}}$, where $e_{b}$ is the bit width of the exponent of the floating point, therefore the expectation of $Z$ is derived as
\begin{equation}
\mathbb{E}\left[Z\right]=\frac{{\cal R}_{\max}+{\cal R}_{\min}}{\left({\cal R}_{\max}-{\cal R}_{\min}\right)}2^{e_{b}-1}\ln{2}\overset{\left(\text{a}\right)}{\approx}2^{e_{b}-1}\ln{2},\label{-203}
\end{equation}
where $\left(\text{a}\right)$ follows that ${\cal R}_{\max}\gg{\cal R}_{\min}$.

The derivation of the mean of $Z^{2}$ is written as
\begin{equation}
\begin{aligned}
\mathbb{E}\left[Z^{2}\right]=\frac{1}{3}\left(\frac{{\cal R}_{\max}}{{\cal R}_{\min}}+\frac{{\cal R}_{\min}}{{\cal R}_{\max}}+1\right)\approx\frac{1}{3}2^{2^{e_{b}}},\label{-4}
\end{aligned}
\end{equation}
highlighting that $\mathbb{E}\left[Z^{2}\right]$ grows exponentially with $e_{b}$.

By substituting \eqref{-203} and \eqref{-4} into \eqref{-205}, we have
\begin{equation}
\mathbb{E}\left[\frac{\partial \sigma_{t,n k}^{2}}{\partial \sigma_{c,(n+1) k^{\prime}}^{2}}\right]=1+\frac{1}{\ln 2} 2^{2-e_{b}}>1,
\end{equation}
referring addition raises the demand for computing precision.

Considering subtraction, similar to addition, we have
\begin{equation}
\mathbb{E}\left[\frac{\partial \sigma_{t,n k}^{2}}{\partial \sigma_{c,(n+1) k^{\prime}}^{2}}\right]=\mathbb{E}\left[\frac{(Z-1)^{2}}{Z^{2}}\right]=1-\frac{2}{\mathbb{E}[Z]}+\frac{1}{\mathbb{E}\left[Z^{2}\right]}.
\end{equation}
The PDF of $Z$ is the same with the one of addition. However, when two operands of subtraction are equal, the relative error of the result tends to infinity and becomes meaningless. Therefore, the domain of $Z$ near 1 needs to be carefully removed. Considering $Z$ is discrete, the domain of $Z$ is
\begin{equation}
Z\in\left[\frac{{\cal R}_{\text{min}}}{{\cal R}_{\text{max}}},1-\frac{2^{-f_{b}}}{1+J2^{-f_{b}}}\right]\cup\left[1+\frac{2^{-f_{b}}}{1+J2^{-f_{b}}}, \frac{{\cal R}_{\text{max}}}{{\cal R}_{\text{min}}}\right],
\end{equation}
where $f_{b}$ is the bit width of the effective number of the floating point, and $J$ is a random variable uniformly distributed within ${0,1,2,\cdots,2^{f_{b}}-1}$. Since the removed domain of $Z$ is sufficiently small, the PDF of $Z$ remains effective. In this way, the expectation of $Z$ is derived in \eqref{-5}, where $\alpha=\frac{2^{-f_{b}}}{1+J2^{-f_{b}}}$. Meanwhile, we have

\begin{figure*}
\begin{equation}
\begin{aligned}
\mathbb{E}[Z] =&\int_{-\infty}^{\infty}z\sum_{j=0}^{2^{f_{b}-1}}\left[f_{Z}\left(z|j\right)\text{P}\left(J=j\right)\right]\text{d}z=\frac{1}{2^{f_{b}-1}}\sum_{j=0}^{2^{f_{b}-1}}\left(\int_{\frac{{\cal R}_{\min}}{{\cal R}_{\max}}}^{1-\alpha}zf_{Z}\left(z|j\right)\text{d}z+\int_{1+\alpha}^{\frac{{\cal R}_{\max}}{{\cal R}_{\min}}}zf_{Z}\left(z|j\right)\text{d}z\right)\\
=&\frac{1}{4\left({\cal R}_{\max}-{\cal R}_{\min}\right)}\bigg\{-\frac{1}{2}\left({\cal R}_{\min}^{2}+{\cal R}_{\max}^{2}\right)+\left({\cal R}_{\min}^{2}-{\cal R}_{\max}^{2}\right)\left(\text{ln}{\cal R}_{\min}-\text{ln}{\cal R}_{\max}\right)+\frac{1}{2^{f_{b}-1}}\sum_{j=0}^{2^{f_{b}-1}}\\
&\bigg[\frac{1}{2}\left({\cal R}_{\min}^{2}+{\cal R}_{\max}^{2}\right)+\left({\cal R}_{\min}^{2}-{\cal R}_{\max}^{2}\right)\alpha+\frac{1}{2}\left({\cal R}_{\min}^{2}+{\cal R}_{\max}^{2}\right)\alpha^{2}-{\cal R}_{\min}^{2}\text{ln}\left(1-\alpha\right)-{\cal R}_{\max}^{2}\text{ln}\left(1+\alpha\right)\bigg]\bigg\}\\\label{-5}
\end{aligned}
\setlength\abovedisplayskip{1pt}
\setlength\belowdisplayskip{1pt}
\end{equation}
\hrulefill
\end{figure*}

\begin{align}
\sum_{j=0}^{2^{f_{b}-1}}\alpha&=\sum_{j=0}^{2^{f_{b}-1}}\frac{2^{-f_{b}}}{1+J2^{-f_{b}}}=\sum_{j=0}^{2^{f_{b}-1}}\frac{1}{j+2^{f_{b}}}\overset{\left(a\right)}{\rightarrow}\ln2,\label{-101}  \\
\sum_{j=0}^{2^{f_{b}-1}}\alpha^{2}&=\sum_{j=0}^{2^{f_{b}-1}}\left(\frac{1}{j+2^{f_{b}}}\right)^{2}\overset{\left(b\right)}{\rightarrow} 0,\label{-102}
\end{align}
where $\left(a\right)$ and $\left(b\right)$ hold true when $f_{b}$ is sufficiently large. By substituting \eqref{-101} and \eqref{-102} into \eqref{-5}, we have
\begin{align}
\mathbb{E}[Z]& \rightarrow \frac{{\cal R}_{\max }+{\cal R}_{\min }}{4\left({\cal R}_{\max }-{\cal R}_{\min }\right)}\left(\ln {\cal R}_{\max }-\ln {\cal R}_{\min }-2 \ln 2\right)\notag\\
&\quad-\frac{{\cal R}_{\max }{ }^{2}+{\cal R}_{\min }{ }^{2}}{4\left({\cal R}_{\max }-{\cal R}_{\min }\right)^{2}}\notag\\
&= 2^{e_{b}-2} \ln 2-\frac{\ln 2}{2}-\frac{1}{4}.\label{-6}
\end{align}

Since $\mathbb{E}\left[Z\right]>0$, we arrive at $\mathbb{E}\left[{\partial \sigma_{t,n k}^{2}}/{\partial \sigma_{c,(n+1) k^{\prime}}^{2}}\right]<1$, which means subtraction can always lower the demand for computing precision.
$\hfill\square$


We conclude the expectation of the number of additions and subtractions needed for a one-bit change in $x^{\text{opt}}$ in Table \ref{tab3} with common configurations applied in the standard of IEEE-754\cite{4610935}. It is worth mentioning that it takes hundreds of additions and subtractions to trigger a one-bit change in computing precision for IEEE-754 double precision. This indicates that the expectation given in Lemma  \ref{theorem of speculation} exerts a marginal impact on optimality, due to the law of large number (LLN).

\begin{table*}
\caption{Expectation of the Number of Additions/Subtractions Needed for One-bit Change in $x^{\rm{opt}}$}
\begin{center}
\begin{tabular}{ccccc}
\toprule
$e_{b}$ & 5 & 8 & 11\\
\midrule
\makecell[c]{Corresponding float point configuration} & \makecell[c]{IEEE-754 half precision} & \makecell[c]{IEEE-754 single precision} & \makecell[c]{IEEE-754 double precision}\\
Addition & 9 & 63  & 493\\
Subtraction & 3 & 30 & 245\\
Addition \& Subtraction & 8 & 114 & 974\\
\bottomrule
\end{tabular}\label{tab3}
\end{center}
\end{table*}

\begin{algorithm}[t]
\label{offline VPC}
\DontPrintSemicolon
  \SetAlgoLined
  \textit{\textbf{Input}} $\alpha$ to specify $G_N$; $\sigma_{N 1}^{2}, \cdots, \sigma_{N k_{N}}^{2}$ with help of $G_{N}$.\;
  Calculate the computing precision for final steps, i.e., $x_{N1}^{\text{opt}},\cdots,x_{Nk_{N}}^{\text{opt}}$ with (9) by deriving $\frac{\partial G_N}{\partial x_{Nk}}=0$, $k=1,\cdots,k_N$.\;
  Establish LUTs for ${G_{\sigma,nk}}$, $G_{o}$ and $x_{nk}^{\text{opt}}$ for arithmetics as $x_{nk}^{\text{opt}}=\text{LUT}_{x^{\text{opt}}}\left({G_{\sigma,nk}}/{G_{o}},u\right)$ using (50) in Appendix F.\;
  \For{$n=N-1,\cdots,1;\,k=1,\cdots,k_{n}$}{
    Derive and log $G_{\sigma,nk}=G_{\sigma,\left(n+1\right)k^{\prime}}\frac{\partial \sigma_{c,(n+1) k^{\prime}}^{2}}{\partial \sigma_{c_{t},n k}^{2}}$, the subscript $(n+1)k^{\prime}$ refers to the result of this arithmetic operation while $nk$ refers to the input of this arithmetic operation.\;
    Derive $x_{nk}^{\text{opt}}=\text{LUT}_{x^{\text{opt}}}\left({G_{\sigma,nk}}/{G_{o}},u\right)$.\;
  }
  \textit{\textbf{Output}} $x_{nk}^{\text{opt}},n=1,\cdots,N-1;\,k=1,\cdots,k_n$.\;
  \caption{Offline VPC}
\end{algorithm}

Algorithm \ref{offline VPC} concludes the proposed offline VPC, where

\begin{align}
\frac{\partial \sigma_{c,(n+1) k^{\prime}}^{2}}{\partial \sigma_{t,n k}^{2}}=\left\{\begin{array}{cl}
1-\frac{1}{\ln 2} 2^{2-e_{b}},&\text{addition} \\
1+\frac{8}{2^{e_{b}} \ln 2-2{\ln 2}-5},&\text{subtraction} \\
1, &\text{multiplication} \\
1, & \text{division} \\ 
\frac{1}{4},&\text{square root}.
\end{array}\right.\label{8}
\end{align}
Note that Algorithm \ref{offline VPC} is generic which only gives the Pareto boundary between average computing bit-width and computing performance (in terms of wireless communication metric, e.g., average sum-rate). Therefore,  Algorithm \ref{offline VPC} does not need utility function $G_{N}$.

With \eqref{8}, we have two remarks as following.

\begin{remark}
In the procedure of AL-VPC, addition and square root demand higher computing precision, multiplication and division maintain the computing precision of the preceding step, while subtraction demands lower computing precision from an expectation standpoint.
\end{remark}
\begin{remark}
The objective of VPC is maintaining the precision of the preceding step rather than the target one. In general, the computing precision tends to decrease progressively.
\end{remark}

\subsection{Online VPC Design}
With the online VPC method, the main idea is to exploit forward propagation instead of backward propagation to obtain the exact value of  ${\partial \sigma_{t,n k}^{2}}/{\partial \sigma_{c,(n+1) k^{\prime}}^{2}}$. This change requires additional steps in the procedure. The general procedure of online VPC is divided into three main steps. We take Fig.~\ref{fig2} as an illustrative example.

\begin{enumerate}[]
\item Obtain the initial computing precision for each arithmetic operation in the first step, i.e., $x_{11}$ and $x_{12}$ in Fig.~2. It is worth noting that only those arithmetic operations whose operands are both initial input numbers need to be considered, as such, $x_{22}$ is not included.
\item Calculate the computing precision for the next arithmetic operation based on the current one, i.e., calculate $x_{21}$ based on $x_{11}$.
\item Determine the final computing precision for the next arithmetic operation. Since $x_{11}$ and $x_{12}$ have calculated their own ``$x_{21}$'', a decision on the final $x_{21}$ needs to be made based on these two results.
\end{enumerate}

The computing precision for all arithmetic operations in the algorithm are obtained by iterating these three steps. In the following parts, we describe them one by one.

\begin{algorithm}[t]
\label{Online VPC}
\DontPrintSemicolon
  \SetAlgoLined
  \textit{\textbf{Input}} $\alpha$ to specify $G_N$; $\sigma_{N 1}^{2}, \cdots, \sigma_{N k_{N}}^{2}$ with help of $G_{N}$.\;
  Calculate the computing precision for final steps, i.e., $x_{N1}^{opt},\cdots,x_{Nk_{N}}^{opt}$ with (9) by deriving $\frac{\partial G_N}{\partial x_{Nk}}=0$, $k=1,\cdots,k_N$.\;
  Establish LUTs for ${G_{\sigma,nk}}$, $G_{o}$ and $x_{nk}^{\text{opt}}$ for arithmetics as $x_{nk}^{\text{opt}}=\text{LUT}_{x^{\text{opt}}}\left({G_{\sigma,nk}}/{G_{o}},u\right)$ using (50) in Appendix F.\;
  \For{each arithmetic operation}{
    \textit{For the simplicity of the form, rename its input as $a$ with $G_{\sigma,a}$, $x_a$, and $b$ with $G_{\sigma,b}$, $x_b$ (if the arithmetic operation is square root, then it only has one input operand $a$). Similarly, denote the output as $c$ with $G_{\sigma,c}$, $x_c$.}\;
    \eIf{$a$ and $b$ are both initial numbers (i.e., there is no $G_{\sigma,a}$ or $G_{\sigma,b}$)}{
        Calculate intermediate $x_{ca}$ and $x_{cb}$ according to $\text{LUT}_{x_{d}}\left(N_{\text{a\&s}}\right)$, then look for $G_{\sigma,a}$ and $G_{\sigma,b}$ using the reverse use of $\text{LUT}_{x^{\text{opt}}}\left({G_{\sigma,nk}}/{G_{o}},u\right)$. (\textit{Step 1})\;
    }
    {
        Calculate $G_{\sigma,a}$ and $G_{\sigma,b}$ using (13), and then look for intermediate $x_{ca}$ and $x_{cb}$ using $\text{LUT}_{x^{\text{opt}}}\left({G_{\sigma,nk}}/{G_{o}},u\right)$. (\textit{Step 2})\;
    }
    \eIf{$x_{ca} \neq x_{cb}$}{
        Calculate $x_c$ using (13) or (14) base on the type of arithmetic operation. (\textit{Step 3})\;
    }
    {$x_c=x_{ca}$}
    Look for $G_{\sigma,c}$ using the reverse use of $\text{LUT}_{x^{\text{opt}}}\left({G_{\sigma,nk}}/{G_{o}},u\right)$.\;
    Execute this arithmetic operation with computing precision of $x_c$.
  }
  \textit{\textbf{Output}} $\mathcal{R}_{Nk}, k=1,\cdots,k_N$ (\textit{the output of the signal processing algorithm})\;
  \caption{Online VPC}
\end{algorithm}

\subsubsection{Step 1: Speculation for initial $x_{1k}^{\rm{opt}}$}
We derive the initial computing precision for each initial arithmetic operation such that the assumption made in Step 2 can be justified.

Since the performance of the algorithm is directly related to $\sigma_{N 1}^{2}, \cdots, \sigma_{N k_{N}}^{2}$, $x_{1k}^{\text{opt}},k=1,2,\cdots,k_{0} $ could be derived by back propagation exploiting the recursion model in \eqref{8} and \eqref{9}. With similar techniques as in offline VPC, we adopt the back propagation method with approximation in Lemma \ref{theorem of speculation} to obtain the initial computing precision.

Exploiting Table~\ref{tab3}, an LUT between the number of additions and subtractions $N_{\text{a\&s}}$ and the bit differences between the computing precision of the first arithmetic and last arithmetic $x_{d}$ is established as $x_{d}=\text{LUT}_{x_{d}}\left(N_{\text{a\&s}}\right)$.

\subsubsection{Step 2: Forward propagation}
Once the computing precision for the previous arithmetic operation been obtained, we calculate the computing precision for the current arithmetic operation based on that. The general procedure is similar to the one in Theorem~\ref{theorem of online vpc}, but in reverse order.

\subsubsection{Step 3: Rule for handling with different $x_{nk}^{\rm{opt}}$s}

If an arithmetic operation involves two operands, each operand will generate its own $x_{nk}^{\text{opt}}$ according to the proposed recursion model in \eqref{8} and \eqref{9}. Denote $G_{\sigma,nk}$ of two operands by $G_{\sigma,nk,1}$ and $G_{\sigma,nk,2}$, then $G_{\sigma,nk}$ is a weighted sum of   $G_{\sigma,nk,1}$ and $G_{\sigma,nk,2}$. For addition and subtraction, it follows
\begin{equation}
G_{\sigma,nk}=\frac{aG_{\sigma,nk,1}+bG_{\sigma,nk,2}}{a+b},
\end{equation}where $a$ and $b$ are two operands, respectively. For multiplication and division, it is
\begin{equation}
G_{\sigma,nk}=\frac{G_{\sigma,nk,1}+G_{\sigma,nk,2}}{2}.
\end{equation}

With the above three steps, we are ready to describe the algorithm of online VPC to solve \eqref{4} in Algorithm \ref{Online VPC}. With a similar method given in Algorithm \ref{offline VPC}, a Pareto boundary of online VPC can also be obtained. Since Algorithm \ref{Online VPC} decides the precision of each arithmetic operations during its execution, an increase in complexity is unavoidable, which is also confirmed theoretically in the following Corollary.

\begin{corollary}\label{theorem of add-on comlexity}
Under some mild conditions, the additional complexity introduced by online VPC is negligible.
\end{corollary}

\noindent\textit{Proof:}
Step 1 start with the calculation of ${\partial \sigma_{t,n k}^{2}}/{\partial \sigma_{c,(n+1) k^{\prime}}^{2}}$. For multiplication, division, and square root, the results are fixed, so there is no added complexity. For addition and subtraction, since the differences between two nearby $G_{\sigma,(n+1) k^{\prime}} / G_{o}$ in $\text{LUT}_{x^{\text{opt}}}$  is $\varepsilon^{2}$ (4 for binary case), the results are approximated using just the exponents of the operands as
\begin{align}
\frac{\partial \sigma_{t,n k}^{2}}{\partial \sigma_{c,(n+1) k^{\prime}}^{2}}&=\frac{a^{2}}{\left(a\pm b\right)^{2}}=\varepsilon^{2\left(\log_{\varepsilon}a-\log_{\varepsilon}\left(a\pm b\right)\right)}\notag\\
&\overset{E_{a}\neq E_{b}}{\approx}\varepsilon^{2\left(E_{a}-\text{max}\left\{E_{a},E_{b}\right\}\right)},
\end{align}
where $E_{a}=\lfloor \log_{\varepsilon}a\rfloor$ is the exponent of $a$, similar for $E_{b}$. It is worth mentioning when $E_{a}=E_{b}$, the approximation is invalid especially for subtraction. When the approximation is valid, $x_{\left(n+1\right)k^{\prime}}^{\text{opt}}$ can be approximated as
\begin{equation}
x_{\left(n+1\right)k^{\prime}}^{\text{opt}}\approx x_{nk}^{\text{opt}}+E_{a}-\text{max}\left\{E_{a},E_{b}\right\}.
\end{equation}
Coincidentally, $E_{a}-\text{max}\left\{E_{a},E_{b}\right\}$ is an existing intermediate result of the realization of arithmetics of addition and subtraction, so the add-on complexity of Step 1 is nearly negligible. 

\begin{figure}
\centering
\includegraphics[width=0.48\textwidth]{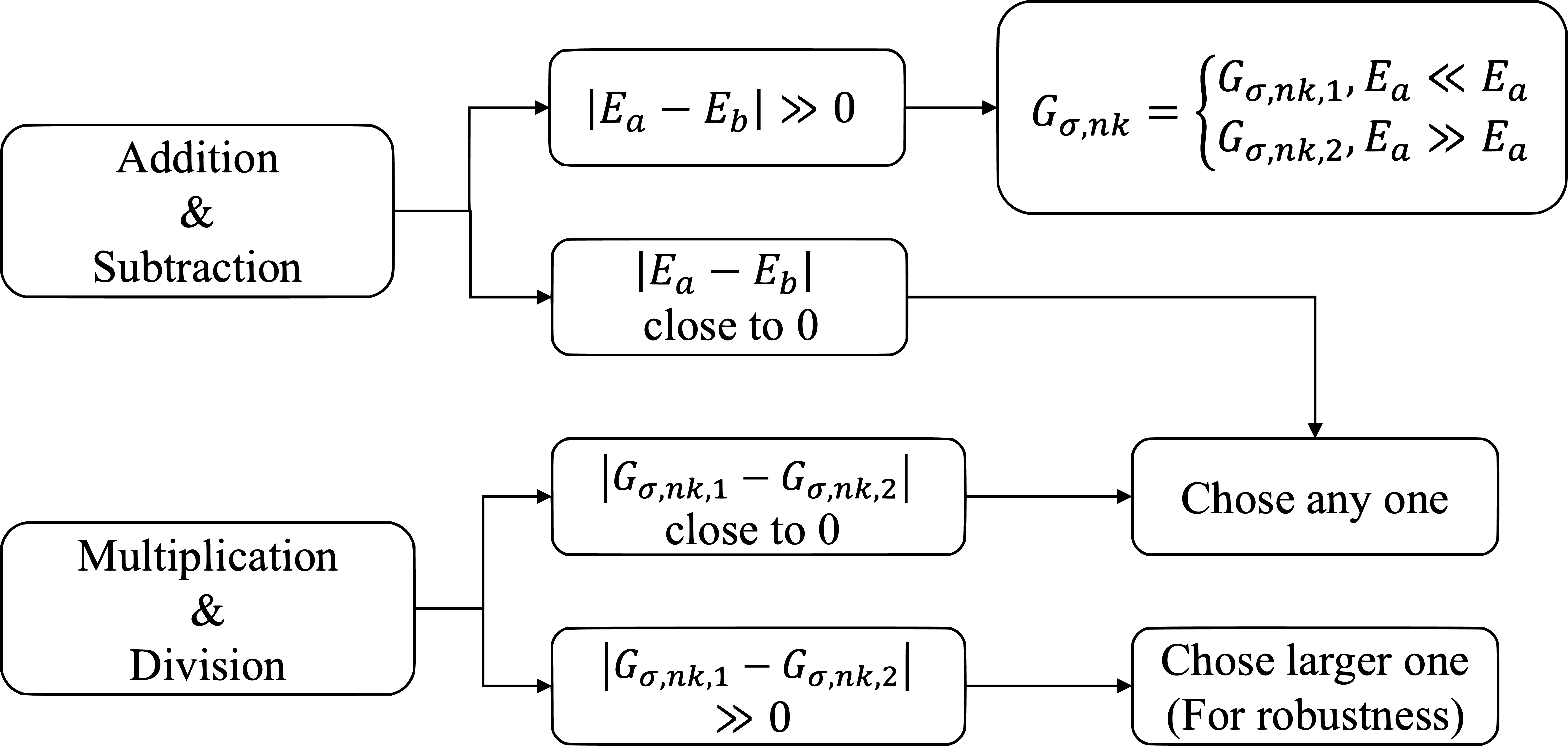}
\caption{The choice of $G_{\sigma,nk}$ in Step 3 for each arithmetic operation.}
\label{add-on}
\end{figure}

For Step 3, though the computation of $G_{\sigma,nk}$ is inevitable, the precision could be set very low. The choice of $G_{\sigma,nk}$ of in Step 3 for each arithmetic operation is shown in Fig. \ref{add-on}. For addition and subtraction, when $E_{a}$ and $E_{b}$ differ greatly, then
\begin{equation}
G_{\sigma,nk}=\left\{\begin{array}{cl}
G_{\sigma,nk,1},&E_{a} \gg E_{b} \\
G_{\sigma,nk,2},&E_{a} \ll E_{b}.
\end{array}\right.
\end{equation}
When $E_{a}$ and $E_{b}$ are close in terms of value, it is appropriate to choose any one or take the arithmetic mean of  $G_{\sigma,nk,1}$ and $G_{\sigma,nk,2}$. For multiplication and division, when $G_{\sigma,nk,1}$ and $G_{\sigma,nk,2}$. significantly differ from each other, which is not common in practice, it is better to assign to the larger one to achieve higher precision for robustness. Overall, the added complexity of Step 3 is also negligible.

To conclude, with proper approximation, the added complexity of Algorithm \ref{Online VPC} is negligible.
$\hfill\square$







In addition, thanks to the forward propagation, Online VPC eliminates the need for the computation graph in advance, which means it supports algorithms with loops and branches. However, in case of algorithms with loops and branches, the initial computing precision can only be arbitrarily determined as it can not be calculated with step 1 anymore. Fortunately, this step primarily influences the precision-complexity trade-off, without affecting the effectiveness of Online VPC.

\section{Case Study on MIMO Precoding Calculation}

In this section, we evaluate the performance of AL-VPC with numerical simulations. Since the optimization in \eqref{4} of AL-VPC involves multiple objectives, without loss of generality, we present the Pareto boundary in terms of computing precision and computational complexity. In the following, we focus on a case study of zero-forcing (ZF) precoding\cite{1261332}. We employ the eBFP storage scheme with $F=1$ to maximum the potential of AL-VPC, $E=10$ to prevent overflow or underflow, and $N=x+1$ which corresponds to the precision of each number. We utilized 100 channel matrices for every point (each computing precision), and these channel matrices remained constant across all points to ensure equivalence between each computing precision. In addition, we only consider the bit width of the mantissa, while the bit width of the exponent is not included.


\subsection{Sum Rate Analysis}

\begin{figure}
\centering
\includegraphics[width=0.5\textwidth]{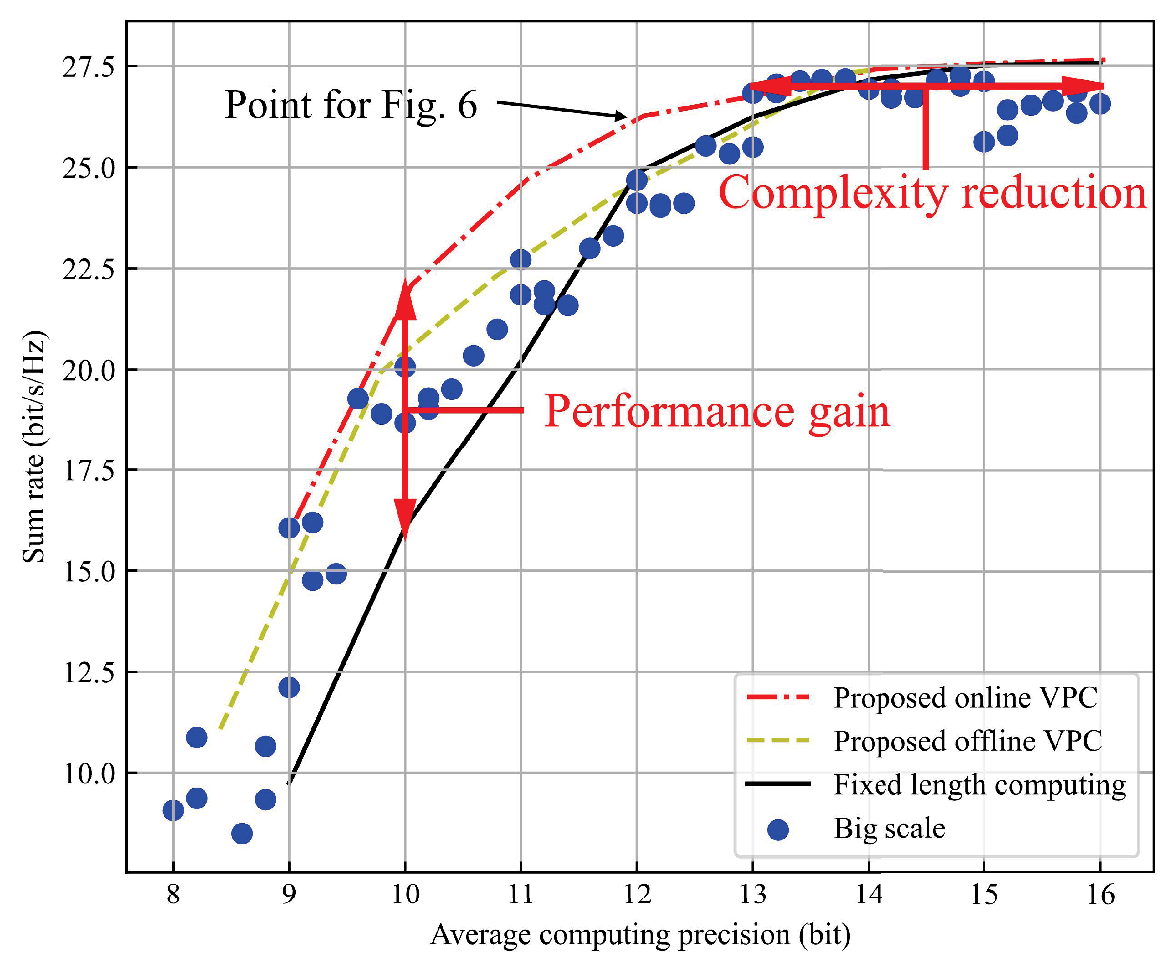}
\caption{Performance and complexity of ZF precoding with AL-VPC, benchmarking against the fixed length computing and random large scale solution.}
\label{fig6}
\end{figure}
\begin{figure}
\centerline{\includegraphics[width=0.46\textwidth]{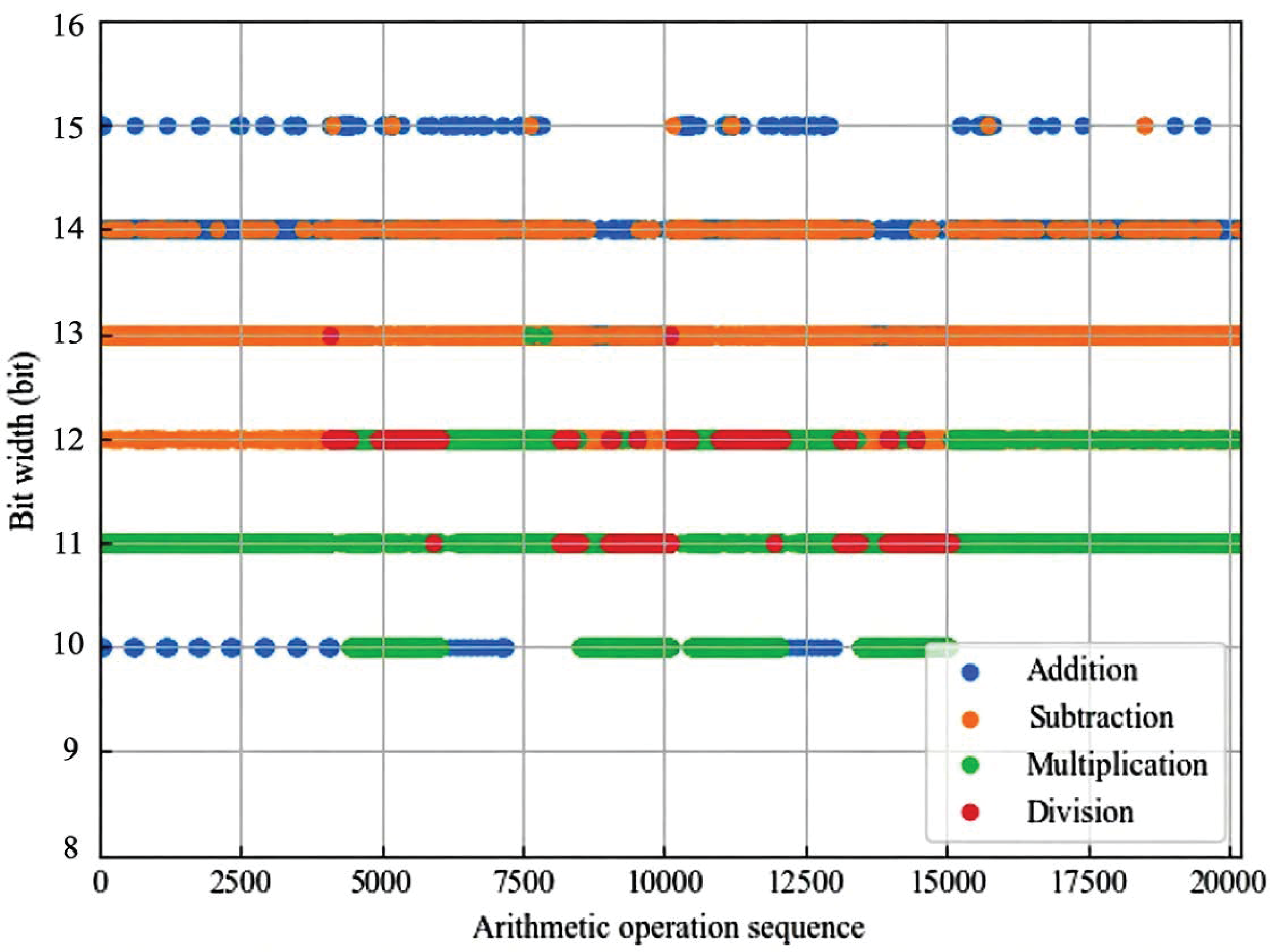}}
\caption{The computing precision of each arithmetic operation of an $8\times8$ ZF precoding.}
\label{precision}
\end{figure}

\begin{figure*}
\centerline{\includegraphics[width=1\textwidth]{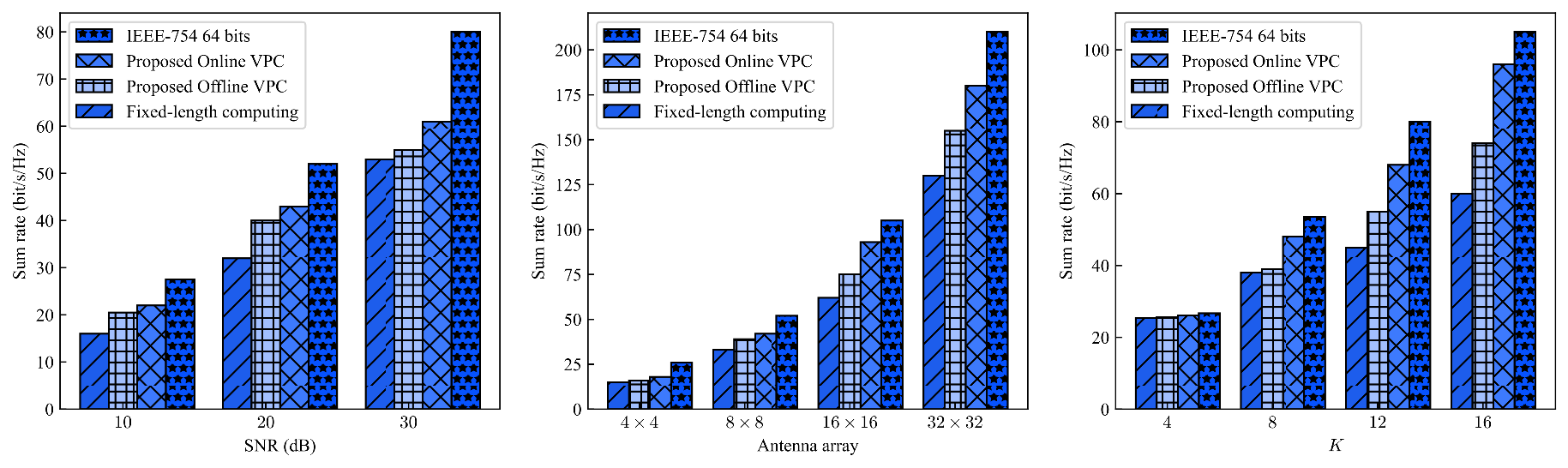}}
\caption{Performance comparison between VPC and fixed length computing in different scenarios.}
\label{fig8}
\end{figure*}
\begin{table*}
\caption{Average Computing Precision Used in Fig.~\ref{fig8}}
\renewcommand\arraystretch{1.5}
\begin{center}
\begin{tabular}{|c|c|c|c|c|}
\hline
SNR figure & SNR 10 dB& SNR 20 dB& SNR 30 dB& /\\ \hline
Average computing precision (bit) & 12 & 14 & 17 &  /\\ \hline
Antenna array figure & $4\times 4$ array & $8\times 8$ array & $16\times 16$ array & $32\times 32$ array \\ \hline
Average computing precision (bit) & 9 & 10 & 12 & 18 \\ \hline
$K$ figure & $K=4$ & $K=8$ & $K=12$ & $K=16$\\ \hline
Average computing precision (bit) & 10 & 10 & 11 & 12 \\ \hline
\end{tabular}\label{acp}
\end{center}
\end{table*}

Fig.~\ref{fig6} considers a communication system with $N_T=8$ transmit antennas and $K=8$ users each with $N_R=1$ receive antenna. The signal-to-noise ratio (SNR), which is defined as the ratio between the total transmit power and channel noise power, is set to 10 dB. The average computing precision is defined as the weighted average computing precision with each basic arithmetic operation. The weight for addition, subtraction, multiplication, division, and square root is 1:1:30:30:80, respectively, based on their time complexity, for both the eBFP design and the IEEE-754 standard. In this case, the value of $G_o$ matches accordingly. The figure illustrates the performance comparison of the sum rate of ZF precoding with different computing schemes, where we also try random large-scale solutions to \eqref{4} by separating the computation of ZF precoding into three parts as
\begin{equation}
\mathbf{W}=\underbrace{\mathbf{H}^{H}\underbrace{(\overbrace{\mathbf{HH}^{H}}^{\circled{1}})^{-1}}_{\circled{2}}}_{\circled{3}}.
\end{equation}

The first and third parts are matrix multiplication, the second part is matrix inversion. Each of these parts is calculated using distinct randomly chosen computing precision. In general, both AL-VPC schemes exhibit superior trade-off capabilities compared to the large-scale solution and fixed-length computing. In addition, the performance gaps between the schemes are more pronounced at lower computing precision levels, as all schemes converge to their best performance at high precision levels. For instance, when considering an average computing precision of 9 bits, the online VPC demonstrates a significant performance gain of nearly 60\% compared to fixed-length computing, while the offline VPC achieves an approximate 50\% improvement. From the perspective of complexity, the online VPC consistently requires approximately 10\% fewer bits for computation than fixed-length computing does. Meanwhile, the offline VPC demonstrates an approximately 8\% reduction in bit usage in the low-precision range. However, this advantage diminishes in the high-precision range.

Fig.~\ref{precision} illustrates the computing precision of each arithmetic operation of one iteration of the point labeled in Fig.~\ref{fig6}. The algorithm consists 20168 basic arithmetic operations with computing precision ranging from 10 bits to 15 bits. Upon closer examination, the computing precision of the majority of arithmetic operations falls within the range from 11 bits to 14 bits, indicating a relatively small range. Furthermore, multiplications tend to utilize relatively low computing precision, whereas additions tend to employ relatively high computing precision. This result corresponds to the Remark~3 in the previous section.



Performance and complexity comparison of the proposed AL-VPC and fixed length computing in different scenarios are presented in Fig.~\ref{fig8} at the top of the next page. For all the subfigures, we deliberately select the computing precision that maximizes the performance gap, and fixed-length computing employs the same average computing precision as the Online and Offline VPC. Specific choices of the average computing precision values are provided in Table~\ref{acp}. Firstly, when considering the influence of SNR,  we set $N_T=8,K=8,$ and $N_R=1$. It is evident that both the online VPC and offline VPC consistently outperform the fixed-length computing. On average, the online VPC achieves a superior performance of 29\% compared to the fixed-length computing, while the offline VPC exhibits an average improvement of 19\%. Secondly, when examining the impact of antenna array scale, we set $\text{SNR}=10\,\text{dB}$, where a larger performance gap is observed between the online VPC, offline VPC, and fixed-length computing. This can be attributed to the fact that the computation steps increase proportionally with the scale of the antenna array. Lastly, when considering the symmetry of the antenna array, we set $N_T=16,\text{SNR}=20\,\text{dB},$ and $N_R=1$. The advantage of the AL-VPC techniques becomes more pronounced as the antenna array becomes more symmetric. This is due to the fact that a more symmetric array requires a larger number of computation steps.


\subsection{BER Comparison}

The BER performance of the proposed AL-VPC is compared with that of fixed-length computing. It is worth noting that AL-VPC is only applied to precoding, while decoding still uses IEEE-754 64 bits for fair comparison.

\begin{figure}
\centerline{\includegraphics[width=0.5\textwidth,height=0.40\textwidth]{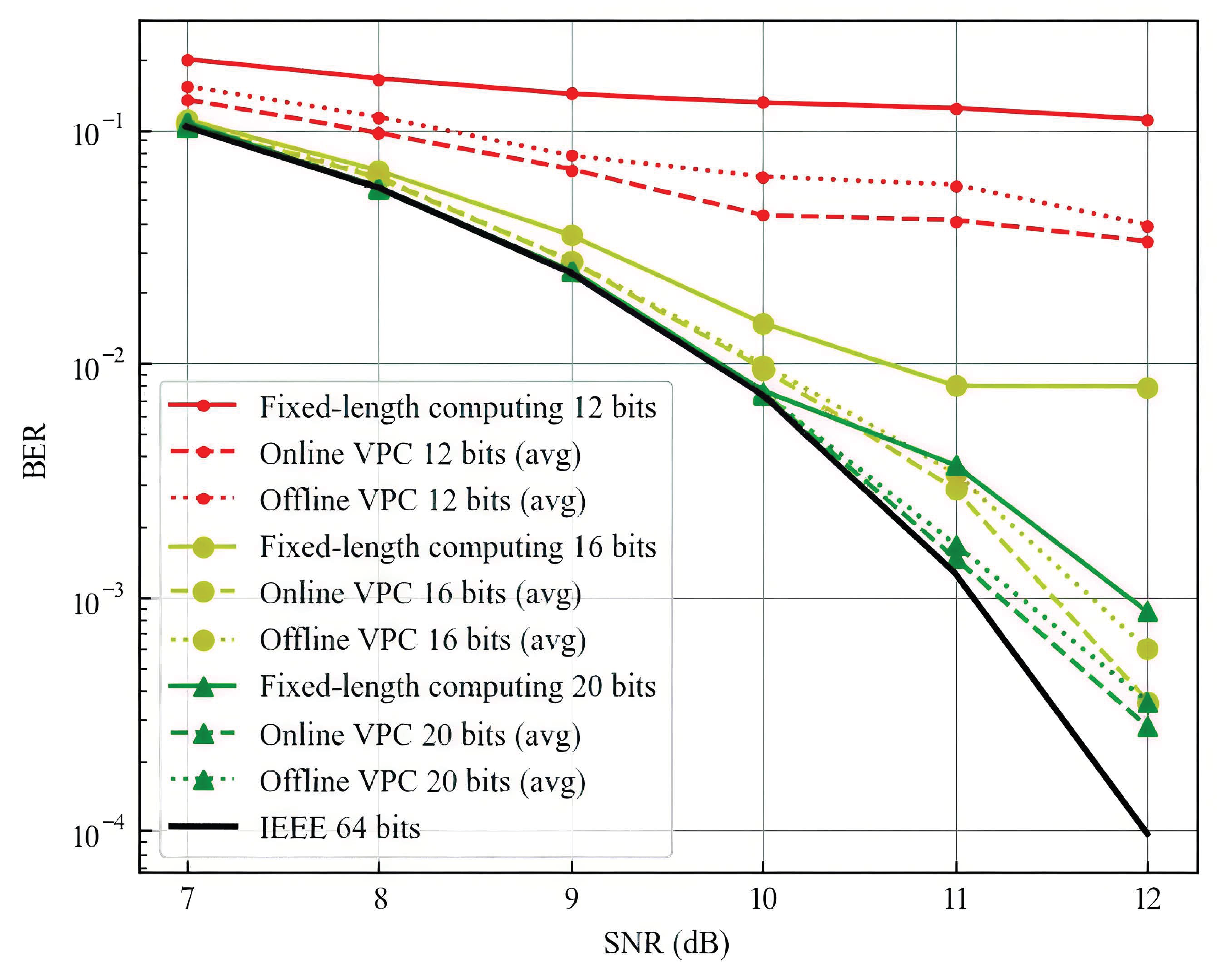}}
\caption{BER performance of ZF precoding with VPC and fixed length computing, $N_T=16$, $K=16$, and $N_R=1$.}
\label{BER_fig1}
\end{figure}
Fig.~\ref{BER_fig1} shows the comparison of BER performance between VPC and fixed-length computing. Regardless of the computing precision, AL-VPC achieves better BER compared to fixed-length computing. When the computing precision is relatively low, although the BER performance of both AL-VPC and fixed-length computing converge as the computing error takes over channel noise, the BER of AL-VPC converges to a lower rate, indicate of AL-VPC's lower computing error.

\section{Conclusion}

In this paper, we investigated the trade-off between computing performance and computational complexity in MIMO signal processing when employing AL-VPC. We proposed the framework of AL-VPC and formulated two optimization problems to trade-off computing performance and complexity. The arithmetic propagation error was first derived to characterize the relationship between the computing precision and the variance of the computing error. Then, we solved the optimization problem with online VPC and offline VPC. Through comprehensive numerical analyses, it becomes evident that the proposed AL-VPC offers a substantial advantage over conventional fixed-length computing, in terms of both computing performance and complexity. To further characterize hardware-specific features of in-memory computing, such as nonlinear variations, the error propagation analysis can be refined accordingly. In addition, the versatile nature of AL-VPC exhibits potential applicability across diverse domains, e.g., MMSE detection and hybrid precoding, that demand heightened performance while adhering to stringent complexity constraints.

\begin{appendices}

\section{An Introduction of eBFP}
\label{eBFP intro}
Existing data storage techniques, e.g., half-precision and single precision arithmetic, present significant limitations that render them unsuitable for AL-VPC adoption. For instance, fixed-point storage suffers from a critical constraint on the range of representable numbers, leading to an increased risk of overflow. Similarly, traditional floating-point storage methods, such as the IEEE-754 standard [1], offer only limited choices that are insufficient to support AL-VPC. This IEEE-754 standard encompasses four options: half precision (16 bits), single precision (32 bits), double precision (64 bits), and quadruple precision (128 bits). The limited range of options causes an inflexible trade-off between computing precision and complexity. To address this issue, it is imperative to design a novel floating-point arithmetic, termed extended block floating-point (eBFP) in our study.

To clarify, it is cruicial to distinguish eBFP from the traditional block floating-point (BFP) systems. Unlike BFP, which involves segmenting the floating bits of the IEEE-754 standard, the proposed eBFP presents a distinct approach to floating-point arithmetic. BFP operates exploiting a block storage scheme aimed at reducing storage complexity by assigning a shared exponent block to numbers of similar scale. This storage structure comprises two components: the exponent block, containing the shared sign and exponent bits for all numbers, and multiple floating blocks, each storing the floating bits associated with a specific number.

\begin{figure}
\centerline{\includegraphics[width=0.48\textwidth]{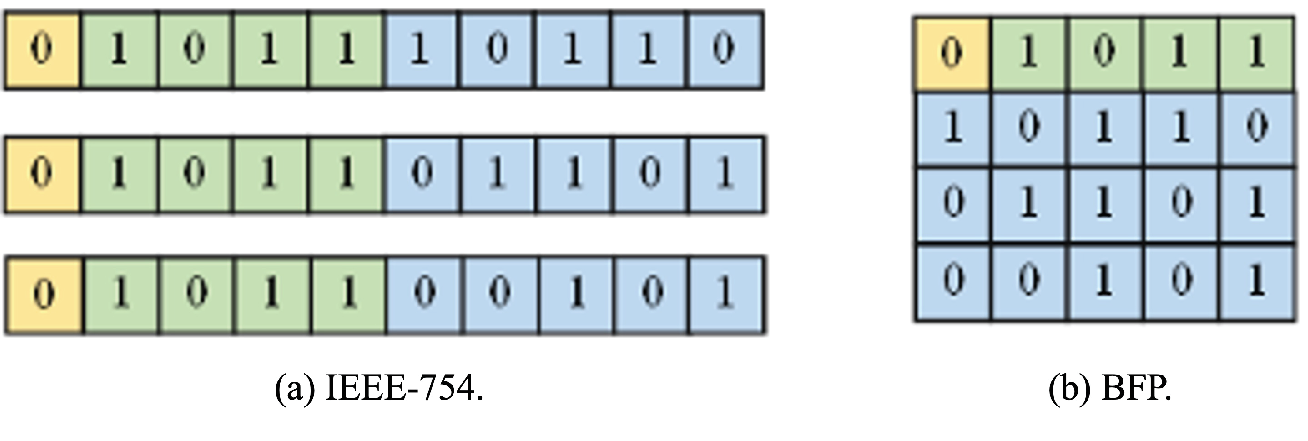}}
\caption{Comparison between IEEE 754 and BFP.}
\label{fig1}
\end{figure}

Fig.~\ref{fig1} presents a schematic comparison between BFP and IEEE-754. For ease of understanding, we adopt a representation with an exponent bit width of 4 and a floating bit width of 5, which is however not explicitly defined in IEEE-754. BFP, as illustrated, extracts common exponent bits and delineates the floating bits for three numbers. It is important to note that BFP does not specify the exact number of exponent and floating bits. Consequently, in practice, the ``shape'' formed by BFP may not necessarily resemble the ``rectangle'' depicted in Fig.~\ref{fig1}, but rather an irregular form. Compared to IEEE-754, which requires 30 bits of storage, BFP utilizes only 20 bits, resulting in significant savings in storage.

Unlike traditional BFP, which stores multiple numbers, each instance of eBFP is designed to accommodate only \textit{a single number}. Instead of storing the exponent corresponding to scientific notation, the exponent bits in eBFP represent the ``serial number'' of the block containing the highest bit. To enhance the flexibility of eBFP, both the number of bits and the sizes of the exponent block and fraction blocks can be adjusted according to specific requirements.

\begin{algorithm}[t]
\renewcommand{\thealgocf}{R.1}
\label{eBFP}
\DontPrintSemicolon
  \SetAlgoLined
  Set the storage range and precision of eBFP with the number of bits of exponent block $E$, the number of bits of fraction blocks $F$, and the number of blocks $N$.\;
  Set the sign bit according to the sign of the decimal, 0 for positive decimal and 1 for negative decimal.\;
  Calculate the exponent of the decimal corresponding to scientific notation, $e_{\text{sci}}=\lfloor\log_2 d\rfloor+1$, where $d$ is the decimal.\;
  Calculate the exponent of eBFP, $e=\lceil e_{\text{sci}}/F \rceil$.\;
  Convert the exponent into IEEE 754 standard, $e_{754}=e+2^{E-2}-1$.\;
  Convert decimal exponent $e_{754}$ into binary one $e_b$.\;
  Combine the sign bit and binary exponent into the exponent block.\;
  Calculate the number of 0s in the first fraction block, $z=e-e_{\text{sci}}/F$.\;
  Convert the high-precision decimal into low-precision binary with $(N-1)F-z$ bits of significant digits with a proper rounding mechanism (e.g., round to the nearest).\;
  Insert 0s in front of the low-precision binary, before cutting into $(N-1)$ fraction blocks.\;
  Insert the exponent block in front of the fraction blocks.\;
  \caption{Convert from High-precision Decimal Number to eBFP}
\end{algorithm}

Algorithm~R.1 outlines the conversion process from high-precision decimal numbers to eBFP. Notably, certain special cases, such as underflow and overflow, are not addressed. To elaborate on the conversion process, we offer a straightforward demonstration. As depicted in Fig.~\ref{demoeBFP}, assuming the number is represented by a fixed point with the highest digit 57 digits before the decimal point, while parameters for the storage range and precision are respectively set as $E=F=8$ and $N=3$. In eBFP, the serial number of the first block (8 bits) to the left of the decimal point is designated as No.~0, with subsequent blocks incrementing to the left and decrementing to the right. Given that the highest bit of the number resides in the block No.~7, the corresponding exponent bit is set to 7. For fraction blocks, the first block is the block where the highest bit resides. If the number of bits in the initial fraction block is fewer than 8, 0's are appended until it reaches 8 bits. Subsequently, based on the specified number of eBFP blocks ($N=3$), blocks with lower serial numbers are sequentially filled. Because the bit after block No.~6 is 0, with round to the nearest, no carry is needed. Therefore the fraction block 1 is block No.~7, and the fraction block 2 is block No.~6.

When setting $E=F=8$ in eBFP, the range of representable numbers expands by 8 times compared to traditional floating-point, achieved by altering the representation of exponent bits. Specifically, 7 bits are utilized to represent a range equivalent to what 10 bits would represent in IEEE-754. This storage range comes close to the storage range (11 bits) covered by IEEE-754 double-precision floating-point (64 bits). However, this method may lead to the presence of a ``0'' in the first floating block. While this ``0'' only encodes the specific position of the highest bit in the block, carrying merely 3 bits of information, it occupies an expected space of 3.5 bits, resulting in a loss of 0.5 bit of information. In addition, IEEE-754 can omit the first bit by employing scientific notation, while eBFP changes the meaning of the exponent bits and cannot omit the first bit. In summary, these three factors suggest that the fraction of eBFP achieves precision flexibility at the cost of 1.5 bits compared to IEEE-754.

Compared to the standard IEEE-754, eBFP maintains a consistent width for the exponent regardless of the number of blocks, ensuring that the range of representable numbers remains unchanged. This feature aligns well with practical needs, as precision control often necessitates maintaining a consistent representation range. With IEEE-754, transitioning to a low-precision representation requires careful consideration of potential underflow and overflow scenarios, posing challenges in designing large numbers with reduced precision. In contrast, \textit{eBFP enables precision adjustments without altering the representation range}. The precision of eBFP is determined by the number of blocks, offering finer control over precision adjustment compared to IEEE-754. Crucially, each eBFP block holds a specific physical significance, facilitating storage and calculations at the block level. In contrast, IEEE-754 lacks such granularity and cannot perform calculations with reduced bit width beyond division into sign bits, exponent bits, and floating bits.

\end{appendices}

\bibliographystyle{IEEEtran}
\bibliography{IEEEabrv, reference}

\end{document}